# Trace Elemental Behavior in the Solar Nebula: Synchrotron X-ray Fluorescence Analyses of CM and CR Chondritic Iron Sulfides and Associated Metal

S. A. Singerling[1*] S. R. Sutton[2, 3], A. Lanzirotti[3], M. Newville[3], and A. J. Brearley[1]

[1]Department of Earth and Planetary Sciences, MSC-03 2040 1 University of New Mexico, Albuquerque, NM 87131, USA
[2]Department of Geophysical Sciences, University of Chicago, Chicago, IL 60637, USA
[3]Center for Advanced Radiation Sources, University of Chicago, Chicago, IL 60637, USA

*Corresponding author email: ssheryl@vt.edu

**Abstract**

We have performed a coordinated focused ion beam (FIB)-scanning and transmission electron microscopy (SEM, TEM), electron probe microanalysis (EMPA)-synchrotron X-ray fluorescence (SXRF) microprobe study to determine phase-specific microstructural characteristics and high-resolution *in situ* trace element concentrations of primary pyrrhotite, pentlandite, and associated metal grains from chondrules in CM2 and CR2 carbonaceous chondrites. This work is the first of its kind to link trace element chemical and microstructural observations in chondritic sulfides in an attempt to determine formation mechanisms and conditions of primary sulfides in these meteorite groups. SXRF and TEM analyses were performed on a small number of FIB sections which act as representative samplesof primary sulfides and associated metal present in pyrrhotite-pentlandite intergrowth (PPI) grains and sulfide-rimmed metal (SRM) grains.

SXRF microprobe analyses allowed the concentrations of the minor and trace elements, Co, Cu, Ge, Zn, and Se to be quantified, in addition to Fe and Ni, at a spatial resolution of 2 μm. The similarity between the CM and CR PPI sulfide trace element patterns provides evidence for a common formation mechanism for this type of sulfide grain in both meteorite groups. In addition, the SRM sulfide and metal have comparable trace element patterns that indicates a genetic relationship between the two, such as sulfidization of metal. Enrichments in Ni, Co, Cu, and Se are consistent with the chalcophile/siderophile behavior of these elements. The observed depletions in Ge suggest that it may have been lost by evaporation or else was never incorporated into the metal or sulfide precursor materials. The depletion in Zn may also be attributable to evaporation, but, being partially lithophile, may also have been preferentially incorporated into silicates during chondrule formation. Trace element concentrations support crystallization from an immiscible sulfide melt in chondrules for formation of the PPI grains and sulfidization of metal for the origin of the SRM grains.

# 1. INTRODUCTION

Recent studies (e.g., Kimura et al., 2011; Harries and Langenhorst, 2013; Schrader et al., 2015; 2016; Singerling and Brearley, 2018) have identified primary iron sulfides, specifically pyrrhotite ($Fe_{1-x}S$) and pentlandite ($(Fe,Ni)_9S_8$), in CM and CR carbonaceous chondrites, which likely formed by different mechanisms, primarily based on textural observations. These primary sulfides occur in several different textural forms, such as pyrrhotite-pentlandite intergrowth (PPI) grains, containing mostly pyrrhotite with lesser amounts of pentlandite, and sulfide-rimmed metal (SRM) grains, consisting of a kamacite core rimmed by pyrrhotite with small amounts of pentlandite (Singerling and Brearley, 2018).

Both groups are hypothesized to have formed in the solar nebula, the PPI grains by crystallization of sulfide melts during chondrule formation event(s) and the SRM grains by sulfidization of Fe,Ni metal by $H_2S$ gas. Trace element data for these two textural groups have the potential to reveal information relating to nebular conditions and processes. Many minor and trace elements in meteorites are highly chalcophile, preferring to be incorporated into sulfide phases. In addition, these chalcophile elements tend to be moderately to highly volatile, having condensed at temperatures <1230 K (McSween and Huss, 2010). Such elements and, by proxy, the phases they are hosted in can shed light on key events in the formational history of the solar system.

In this study, we have measured trace element concentrations in primary iron sulfides (pyrrhotite and pentlandite) and associated Fe,Ni metal in CM and CR chondrites. The results provide constraints on their formation mechanisms and improve our understanding of the partitioning behavior of these elements in sulfides. Currently, there are very few analyses of trace elements in meteoritic sulfides and none for individual phases in CM and CR sulfides/metals.

The vast majority of past trace element studies of extraterrestrial materials involved bulk whole rock analyses (INAA, RNAA) (e.g., Grossman and Wasson, 1985; Kallemeyn et al., 1989; Wasson and Kallemeyn, 1988). However, many cosmochemical and geochemical problems require the determination of trace element concentrations in individual phases such as, for example, studies of solid-liquid or solid-solid partitioning (Paktunc et al., 1990; Fleet et al., 1999; Helmy et al., 2010) or trace element zoning within sulfide grains or throughout an entire ore deposit (Barnes et al., 2006; Large et al., 2009; Dare et al., 2012). Sulfides in chondritic meteorites, especially in carbonaceous chondrites, often have grain sizes <50 µm. In such cases, *in situ* analyses using microanalytical techniques, such as Secondary Ionization Mass Spectrometry (SIMS), Laser Ablation Inductively Coupled Plasma Mass Spectrometry (LA-ICP-MS), or SXRF, are necessary. Trace element concentrations in terrestrial sulfides have been determined using SIMS with a spot size of a few microns and minimum detection limits down to the ppb level (e.g., Cabri and McMahon, 1995; Steele et al., 2000). A broad suite of elements has also been determined using LA-ICP-MS with a spatial resolution down to a micron and minimum detection limits of a few ppb (e.g., Barnes et al., 2006; Large et al., 2009; van Acken et al., 2012). Of interest to the current study are LA-ICP-MS analyses of metal and sulfides. Several studies of chondritic metals have been performed using this technique (e.g., Campbell et al., 2002; Humayun et al., 2002; Jacquet et al., 2013; Andronikov et al., 2015; Weyrauch et al., 2021); however, analyses of chondritic sulfides from CM and CR chondrites are lacking. Additionally, SIMS and LA-ICP-MS are both destructive techniques; the former sputters analysis regions, whereas the latter generates laser ablation pits. Such destructive techniques limit the ability to perform microstructural work after the trace element analyses, which is one of the main aims of the current study.

For high spatial resolution trace element analyses, SXRF microprobe is superior to SIMS and LA-ICP-MS, owing to the fact that it is a non-destructive technique and can have submicron level (e.g., 100 nm) spatial resolutions, depending on the beamline (Sutton et al., 2002). For sulfide analyses, the detection limits of SXRF microprobe are comparable, if not superior to these other techniques, and are typically in the ppm to ppb range. SXRF has been used previously to measure trace elements in sulfides in extraterrestrial samples, including troilite (FeS) in lunar and Martian basalts (Papike et al., 2011); pyrrhotite in interplanetary dust particles (Flynn et al., 2000); pyrrhotite in the CI chondrite Orgueil (Greshake et al., 1998); troilite, pentlandite, and metal in CV chondrites (Brearley, 2007); and X-ray mapping of sulfides in the CM2 chondrite Murchison (Dyl et al., 2014). However, with the exception of Greshake et al. (1998) and Flynn et al. (2000), these studies have been carried out on polished thin sections. This results in large sample excitation volumes inherent in the fact that, depending on the material being analyzed and the energy of the fluorescence line being detected, the X-ray beam can excite fluorescence through the entire thickness of the thin section (typically 30 μm). This further degrades the spatial resolution of the analyses compared to the optimum ~μm level.

In addition, the small grain sizes of primary iron sulfides in CM and CR chondrites and the presence of exsolution textures means that analyses performed on polished thin sections may not sample only the region of interest, because the grain may not extend all the way through the thin section. Consequently, the analytical beam may interact with other materials below the sample surface. To avoid the unintentional analysis of undesired phases and maximize spatial resolution, we utilize a newly-developed technique involving SXRF analyses on focused ion beam (FIB)-prepared sections, similar to the methods outlined in Cook et al. (2015). Additionally, using FIB sections allows us to perform TEM analyses on the same material analyzed for trace elements, which links trace element compositional data with microstructural information. This work continues the advances in coordinated analyses of extraterrestrial materials and further increases our knowledge and understanding of the formation and subsequent evolution of these samples.

## 2. METHODS

All analytical work, with the exception of SXRF analyses, was performed at the University of New Mexico (UNM) in the Department of Earth and Planetary Sciences and Institute of Meteoritics. The samples studied were the least-altered carbonaceous chondrites, CM2 QUE 97990,31, CR2 QUE 99177,19, and CR2 EET 92042,35, which were obtained from the NASA Astromaterials Acquisition and Curation Office. Initial characterization was performed on polished thin sections of these meteorites. For back-scattered electron (BSE) imaging of the grains and energy dispersive X-ray spectrometry (EDS) analysis, we used a FEI Quanta 3D DualBeam® Field Emission Gun Scanning Electron Microscope (FEGSEM). EDS analyses were collected with the Genesis EDS system coupled to an EDAX Apollo 40 $mm^2$ silicon drift detector (SDD). The FEGSEM operated under the following conditions for data collection: 10 mm working distance, 10 kV (BSE imaging) or 30 kV (EDS analyses and maps) accelerating voltage, 16 nA beam current. The imaging conditions (i.e., 10 kV) resulted in a higher spatial resolution and were optimal for observing the very fine-scale textures present in many of the grains.

Preparation of focused ion beam (FIB) sections for SXRF microprobe analyses and Transmission Electron Microscope (TEM) studies were also performed on the same FEI Quanta 3D Dualbeam® FEGSEM/FIB instrument. The FIB-prepared samples were removed from the thin section using the *in situ* lift out technique with an Omniprobe 200 micromanipulator and mounted onto molybdenum TEM half grids. We attempted to use silicon TEM half grids initially, but the brittle nature of Si caused breakage of the grid during routine handling both with tweezers and during micromanipulation. Extraction of the FIB samples was performed at an ion beam accelerating voltage of 30 kV with a beam current ranging between 1 and 5 nA. The FIB sections were mounted on the side of the Mo grid's spline to provide the correct geometry for the SXRF analyses (Fig. 1). Molybdenum was chosen for the grid material because any fluorescence generated by scattered radiation would fall outside the energy range of interest. Our preliminary developmental work, to determine optimum parameters of the FIB sections for SXRF analyses, showed that excellent spectra with low backgrounds could be obtained from FIB samples that were ~1–2 µm in thickness.

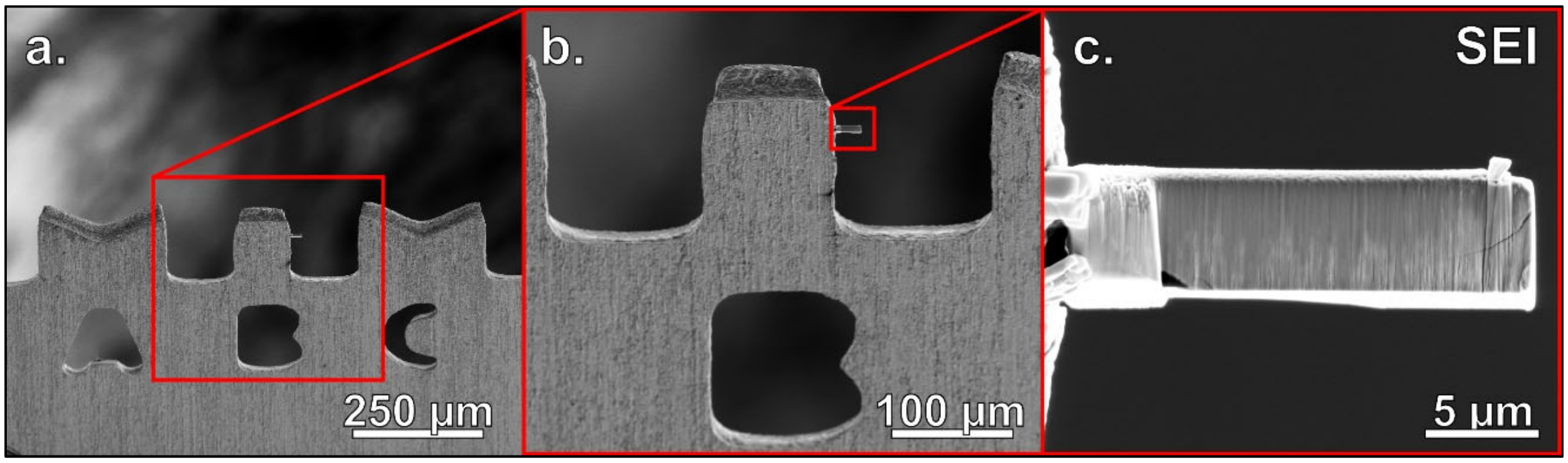

**Figure 1.** Secondary electron SEM images of (a) the Mo TEM half grid, (b) the spline where the FIB section is mounted for SXRF analyses, and (c) the FIB section of PPI grain pyrrhotite from CM QUE 97990.

Final thinning to electron transparency for TEM analyses (~100 nm), performed on just a portion of each FIB section rather than the entire section, was carried out at an ion beam accelerating voltage of 30 kV, with beam currents decreasing from 0.5 nA to 50 pA at the final stage of ion thinning. Platinum straps are usually applied to FIB sections in order to protect the sample from ion beam damage during final thinning to TEM thickness (i.e., electron transparency). However, preliminary SXRF work to determine the optimum parameters for analyses of FIB sections, indicated a large interference from Pt due to the presence of the Pt strap. With this consideration in mind, the FIB sections prepared for this study did not have a Pt strap applied until after the SXRF analyses were performed and final thinning down to electron transparency for TEM analyses was required. Nevertheless, the FIB sample preparation technique still introduces several trace element contaminants into the samples, even without Pt coating. These include Pt that is used to weld the FIB section to the Mo grid, Ga implanted into the sample from the ion beam, and W from the Omniprobe needle used for micromanipulation. We detected spectral artifacts at the following energies from these elements: Pt (9.44, 10.85, 11.07, 11.25 keV), Ga (9.24, 10.26 keV), and W (8.40, 9.67, 9.96 keV). These contaminants overlap with and, consequently, prevent analyses of the following elements of interest for sulfides and metals: Ga, W, Re, Os, Ir, Pt, and Hg.

The major and minor element compositions of the sulfides (Fe, S, Ni, Co, and Cr) were obtained on the polished thin section of the sample prior to FIB section extraction, using wavelength dispersive spectrometry (WDS) on a JEOL 8200 Electron Probe Microanalyzer (EPMA). These data are included in Table A1. Operating conditions were 15 kV accelerating voltage, 20 nA beam current, and a beam size of <1 μm. Elements analyzed, the crystals they were measured on, count times, detection limits, and standards used are summarized in Table A2. Sulfides and metals are listed separately as they were collected using different settings and calibration files. Appropriate corrections were made for elements whose peaks interfere with one other (i.e., Fe and Co) using the Probe for EPMA software (Donovan et al., 1993). We applied standard ZAF corrections within Probe for EPMA to our compositional data.

Minor (Ni and Co) and trace element analyses of the phases (Cu, Zn, Ge, and Se) were obtained using the SXRF Microprobe at the GSECARS beamline (13-ID-E) at the Advanced Photon Source (Argonne National Laboratory). All data related to these analyses is summarized in Table A3. For our discussion and presentation of Co concentrations, we opted to use EPMA rather than SXRF data, owing to the overlap between the Fe Kβ and Co Kα peaks. The higher resolution of WDS analyses from EPMA, as well as a function within the Probe for EPMA software, allowed us to correct for this interference. The SXRF Co data are mostly artificially high because of this overlap, and likely do not represent the true Co concentrations of the sulfides and metals. The use of FIB sections with the SXRF microprobe allowed us to avoid issues with beam overlap of adjacent phases, achieve high spatial resolutions (~2 μm) with minimum detection limits from 1 ppb to 2.9 ppm (Table A3), depending on the element and the phase, and analyze a significantly lower excitation volume compared to polished thin section analyses. The SXRF analyses were carried out on FIB sections of grains that were previously characterized by SEM and EPMA. Four to five spectra were collected on different regions of each of the six FIB section using spot analyses with the beam energy above the absorption edges of interest (i.e., 19 keV) with the detector heavily filtered with aluminum to suppress Fe K fluorescence.

The basic procedure for obtaining quantitative compositional information from the SXRF spectra was to fit the background and fluorescence peaks, subtract the background to obtain net peak areas, and then use these net areas to compute concentrations. Since the background is a smoothly varying function of energy, it could be approximated with sufficient accuracy using spline fitting of spectral regions lacking other features. Fitting of the various fluorescence peaks is generally straightforward because the energies of fluorescence lines are well known and information on the transition probability ratios can be utilized to constrain relative peak intensities (e.g., Fe $K_\alpha$/Fe $K_\beta$). The energies of potentially significant pile-up and escape peaks are also well-defined.

Quantification of the elements in terms of concentration was obtained from the net peak areas using a standardless method based on the *NRLXRF* program (Criss et al., 1978). Iron was used as an internal sensitivity reference using Fe concentrations determined by EPMA for standardization. In this approach, the concentration $C_X$ of a trace element *X* is given by:

$$C_X = C_{Fe} \times \left(\frac{I_X}{I_{Fe}}\right) \times \left(\frac{S_{Fe}}{S_X}\right)$$

where $C_{Fe}$ is the concentration of Fe in the grain determined by EMPA, $I_X$ and $I_{Fe}$ are the peak areas of element X and Fe, respectively, and $S_X$ and $S_{Fe}$ are the element sensitivities (counts/sec/mg $kg^{-1}$) for the two elements. Only the sensitivity *ratio* is required here which greatly improves the accuracy of the quantification. The prediction mode of *NRLXRF* generates relative fluorescence yield values including corrections for secondary fluorescence, thickness, major element composition, density, and analysis geometry.

A JEOL 2010F FASTEM Field Emission Gun Scanning Transmission Electron Microscope (FEGSTEM), operating at 200 kV, was used to obtain high-angle annular dark-field (HAADF) STEM images and selected area electron diffraction (SAED) patterns of the FIB-prepared samples, after final thinning. *In situ* X-ray analyses, both spot and EDS maps, were obtained with an Oxford Instruments AZtecEnergy EDS system coupled to an Oxford X-Max 80N 80 $mm^2$ SDD. The TEM studies were performed on these FIB sections after SXRF analyses in an attempt to determine the homogeneity of the phases on a submicron scale. Unfortunately, three of the six FIB sections, both the pyrrhotite and pentlandite FIB sections from the CM QUE 97990 PPI grain and the SRM metal grain from CR EET 92042, were lost between the SXRF analyses and the final stage of ion thinning for TEM studies was completed. Coordinated TEM studies were performed on the remaining three FIB sections, specifically the PPI pyrrhotite and pentlandite FIB sections from CR QUE 99177 and SRM pyrrhotite from CR EET 92042.

To summarize the flow of our analytical laboratory work, we first imaged the grains of interest with the SEM, obtained major and minor elemental abundances with the EPMA, prepared ~1 μm thick FIB sections, obtained minor and trace elemental abundances of the FIB sections with the SXRF microprobe, thinned the FIB sections down to electron transparency, and analyzed the FIB sections for microstructural features and compositional heterogeneities with the TEM. Given the time-consuming nature of the work, we were only able to perform such coordinated analyses on a limited set of samples. Therefore, we opted to study the dominant phases (pyrrhotite, pentlandite, metal) from the major textural groups (PPI, SRM) observed in the CM and CR chondrites from our previous work (Singerling and Brearley, 2018). We argue that the grains chosen for this study are representative of those from our previous work based on their textural and compositional features. The PPI grains contained pyrrhotite-pentlandite exsolution textures, such as patches, blades, and rods; whereas the SRM grain was characterized by low-Ni metal rimmed with pyrrhotite. We also aimed to study grains that were large enough to allow extraction of FIB sections that were free of compositional heterogeneities on the scale visible with resolution of SEM and EPMA techniques. The EPMA data in Table A1 show quite homogeneous metal, pyrrhotite, and pentlandite, as illustrated by the small range in compositions between the different spot analyses. In Figure A1, we compare the abundances from the EPMA analyses (Co, Ni, Fe, and S) of the grains featured in this study to values in the literature (Singerling and Brearley, 2018). Although there is variation, we argue that the grains chosen are representative of the overall populations of the textural groups (i.e., CM PPI, CR PPI, and CR SRM).

# 3. RESULTS

## 3.1. Textures

Details of the physical and textural characteristics of the grains from which the FIB sections were extracted are provided in Table 1. Figure 2 shows details of the occurrence of the grains and their textural features. As shown in Figures 2b, d, and f, two FIB sections were extracted from different regions of the same grain for three different grains. Each section within a given grain sampled adjacent regions of pyrrhotite or pentlandite (b and d) for the PPI grains and pyrrhotite or metal (f) for the SRM grain. This allowed for the study of elemental partitioning between two coexisting phases. This approach is not possible using conventional SXRF on polished thin sections for these sulfide grains owing to the complex microscale intergrowth of the phases. Additionally, the regions of pentlandite in the grains selected for FIB section extraction have average diameters of no more than 20 µm and 8 µm in the CM and CR chondrites, respectively. The grains from which FIB sections were extracted include a PPI grain in CM QUE 97990 (Fig. 2a–b), a PPI grain in CR QUE 99177 (2c–d), and a SRM grain in CR EET 92042 (2e–f).

**Table 1.** Physical and textural characteristics of the FIB sections, and the grains from which they were extracted, analyzed in this study.

| | Sample | Textural Group | Spatial Occurrence | Phase | Shorthand Notation | FIB Section Thickness (µm)[1] | Lost Prior to TEM Analysis |
|---|---|---|---|---|---|---|---|
| CM | QUE 97990 | PPI | Type IIA chondrule | Po | CM-PPI-Po | 1.11 | Yes |
| | | | | Pn | CM-PPI-Pn | 1.35 | Yes |
| CR | QUE 99177 | PPI | Type IIA chondrule | Po | CR-PPI-Po | 0.94 | No |
| | | | | Pn | CR-PPI-Pn | 1.28 | No |
| CR | EET 92042 | SRM | Type I chondrule | Po | CR-SRM-Po | 1.32 | No |
| | | | | M | CR-SRM-M | 1.44 | Yes |

PPI = pyrrhotite-pentlandite intergrowth, SRM = sulfide-rimmed metal, po = pyrrhotite, pn = pentlandite, m = metal.
[1]Thickness of FIB section for SXRF analyses. The sections were thinned down to electron transparency (~100 nm) after SXRF analyses and prior to TEM analyses.

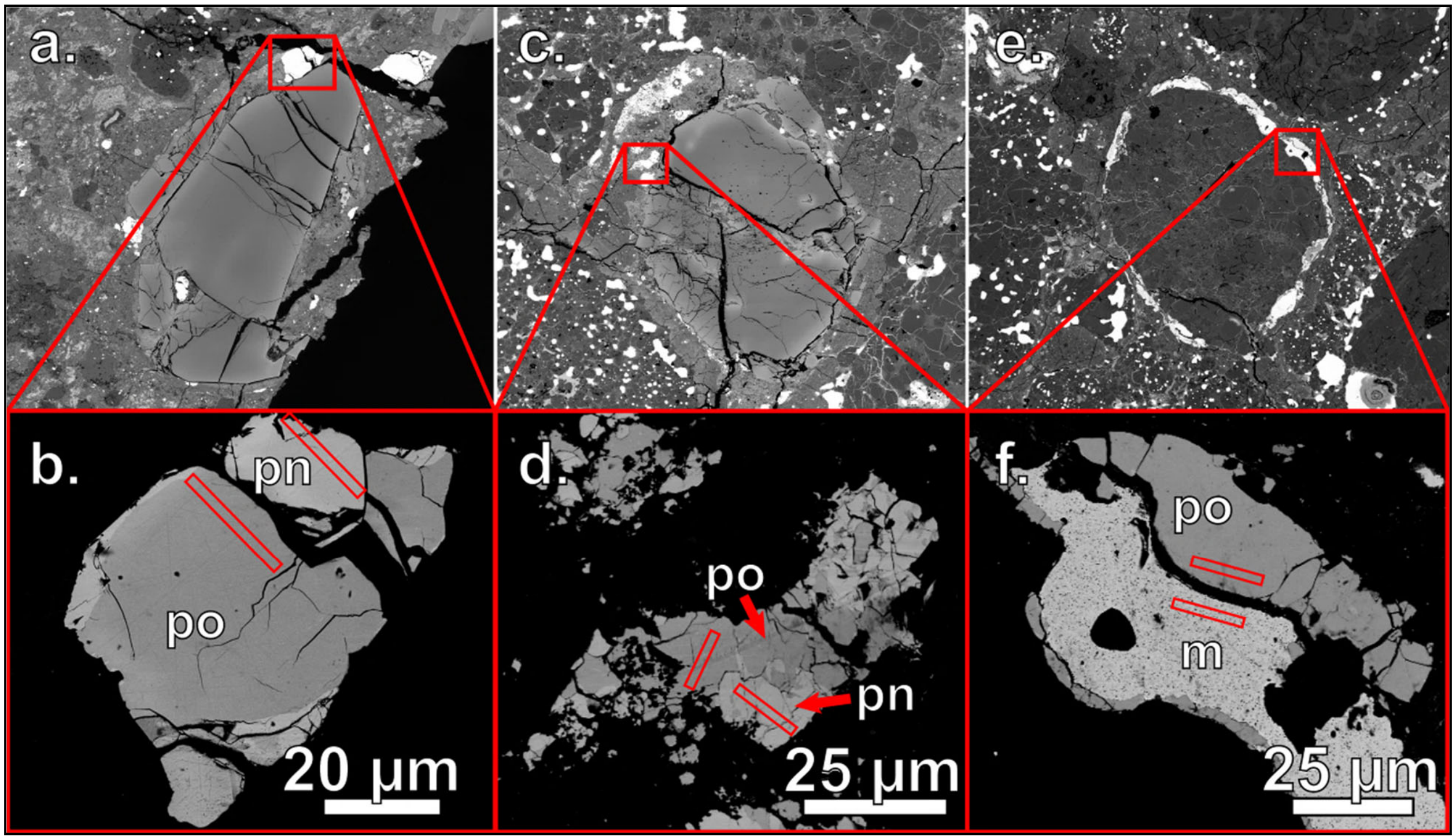

**Figure 2.** BSE images of (a–b) a PPI grain in a type IIA chondrule from CM QUE 97990, (c–d), a PPI grain in a type IIA chondrule from CR QUE 99177, and (e–f) a SRM grain in a type IA chondrule from CR EET 92042, with the locations where pyrrhotite and pentlandite FIB sections were extracted (red rectangles). po = pyrrhotite, pn = pentlandite, m = Fe,Ni metal.

Throughout the remainder of the paper, we will refer to observations and analyses from these FIB sections using shorthand notations for either the meteorite group and specific sulfide phase or textural group and sulfide phase, as summarized for each grain in Table 1. These different grains were chosen as being representative of their textural groups, based on their textural homogeneity at the resolution of the FEGSEM. Additionally, they do not show evidence of aqueous alteration, such as the presence of porosity, reaction fronts, or alteration to different phases (Singerling and Brearley, 2020). All SRM grains in CM chondrites observed in our previous work (Singerling, 2018) show some degree of alteration; hence, an SRM grain from a CM chondrite was not included in this study. The PPI grains in CM and CR chondrites occur predominantly in type IIA chondrules and the matrix, though they have been observed in type IA chondrules as well. These grains likely formed by crystallization of sulfide melts in chondrules during the chondrule formation event(s) (Singerling and Brearley, 2018). The SRM grains in CR chondrites occur in type IA and, less often, type IIA chondrules, but are not observed in the matrix. These grains likely formed by sulfidization of Fe,Ni metal by $H_2S$ gas in the solar nebula (Singerling and Brearley, 2018).

TEM EDS X-ray maps and HAADF STEM images of three FIB sections that survived transport and handling for SXRF analysis are shown in Figures 3 and 4. While compositional variations are difficult to discern in the HAADF images (Fig. 3b, 3e, 4b), they are clearly apparent in the EDS X-ray maps (Fig. 3c, 3f, 4c–d). Submicron pentlandite inclusions (green dots/rods in Fig. 3f and red dots/rods in Fig. 4c–d) are common in the pyrrhotite sections from both the PPI and the SRM grains. They range in shape from round to rod-like and in size from 90 to 790 nm in PPI pyrrhotite and from 10 to 590 nm in SRM pyrrhotite. They are heterogeneously distributed throughout the sections, but are crystallographically oriented with respect to the host pyrrhotite in both the PPI grain and the SRM grain. In the PPI grain, the relationship, as determined with SAED, is $(110)_{pn}//(\bar{1}15)_{po}$, whereas in the SRM grain, the relationship is $(110)_{pn}//(\bar{1}00)_{po}$. Manganese-bearing inclusions (red dots in Fig. 3f and blue dots in Fig. 4c–d) are often found in association with the pentlandite inclusions, occurring at the ends of rods or adjacent to round pentlandite inclusions, though they are far less common than the pentlandite inclusions. They are round in shape and range in size from 70 to 180 nm in PPI pyrrhotite and from 40 to 100 nm in SRM pyrrhotite. Chromium-bearing inclusions (blue dots in Fig. 3f) are also present, but are the least common type of inclusion, and only occur in the PPI grain. They are round in shape and range in size from 60 to 180 nm.

In terms of their microstructures, pyrrhotite from the PPI and SRM grains appear quite similar. However, the sole pentlandite section, from the CR PPI grain, shows heterogeneity in the form of pyrrhotite and a silicon-rich phase. Figure 3c shows a vein-like pyrrhotite feature (in green) and an unidentified silicon-rich phase (in blue) surrounded by the pyrrhotite. With exception of the pyrrhotite and Si-rich phase, the pentlandite appears to be free of other phases and is more homogeneous than either PPI or SRM pyrrhotite, consistent with SEM observations.

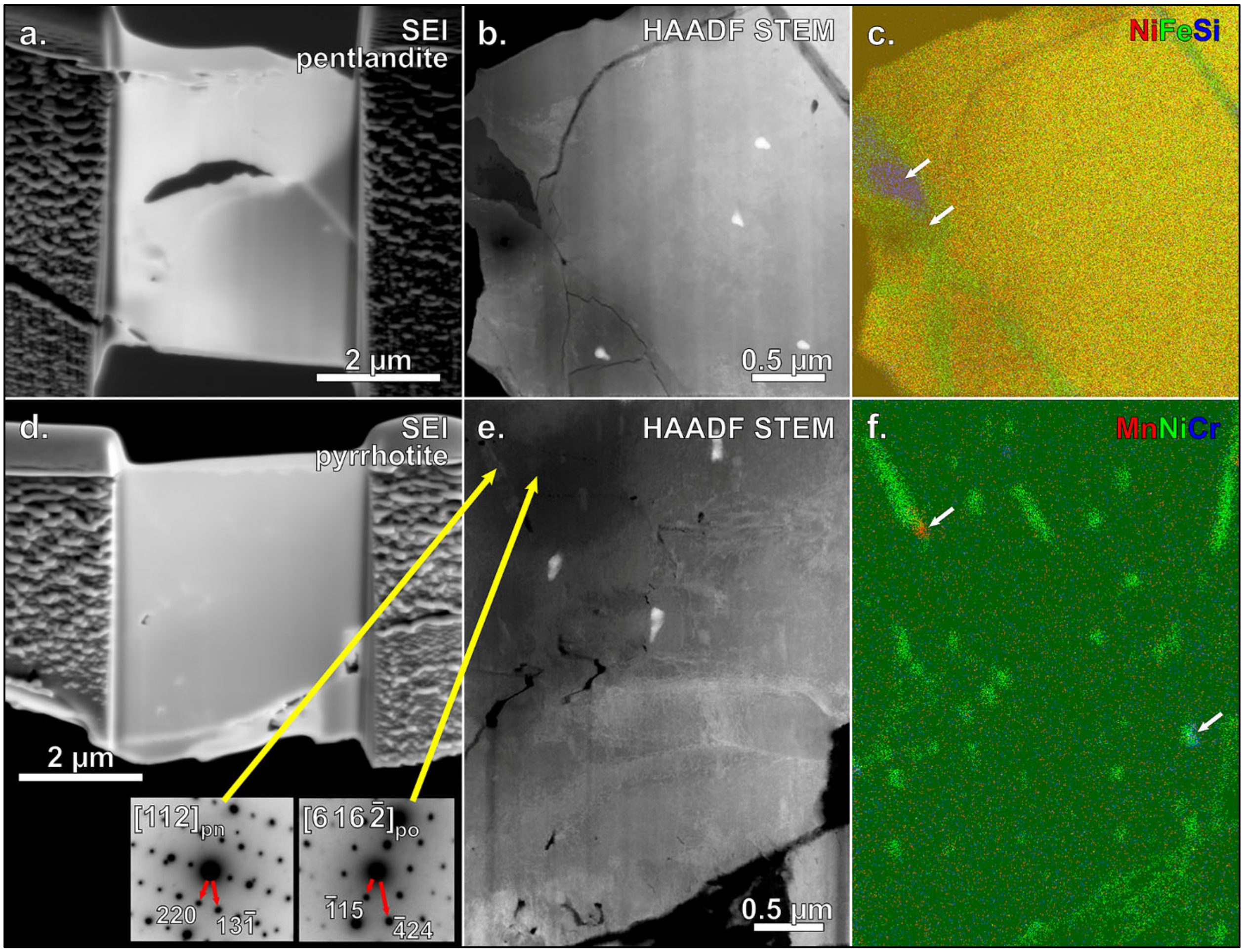

**Figure 3.** SEM and TEM images of a PPI grain located in a type IIA chondrule of CR QUE 99177. BSE images showing the textural context of the grain and the grain itself, along with the location of the FIB section extractions, are presented in Figure 2c and 2d, respectively. A secondary electron image (SEI) (a) shows the full FIB section of the pentlandite section with small regions in the center thinned to electron transparency. A HAADF STEM image (b) showing a higher magnification image of the thinned portion. Electron diffraction patterns are also included from a pentlandite inclusion and the pyrrhotite with their orientation relationships shown. A composite STEM EDS X-ray map of Ni (red), Fe (green), and Si (blue) (c) showing the presence of pyrrhotite (in green), as well as a Si-rich region (in blue), both indicated by white arrows. An SEI (d) showing the overall FIB section of the pyrrhotite section with small regions in the center thinned to electron transparency. A HAADF STEM image (e) showing the thinned portion at higher magnification with pentlandite visible as small, bright inclusions. A composite STEM EDS X-ray map of Mn (red), Ni (green), and Cr (blue) showing the presence of pentlandite inclusions, as well as Mn-bearing (in red) and Cr,Ni-bearing (in teal) inclusions, the latter two emphasized by the white arrows. The white spots in (b) and (e) are carbon contamination marks from EDS analyses.

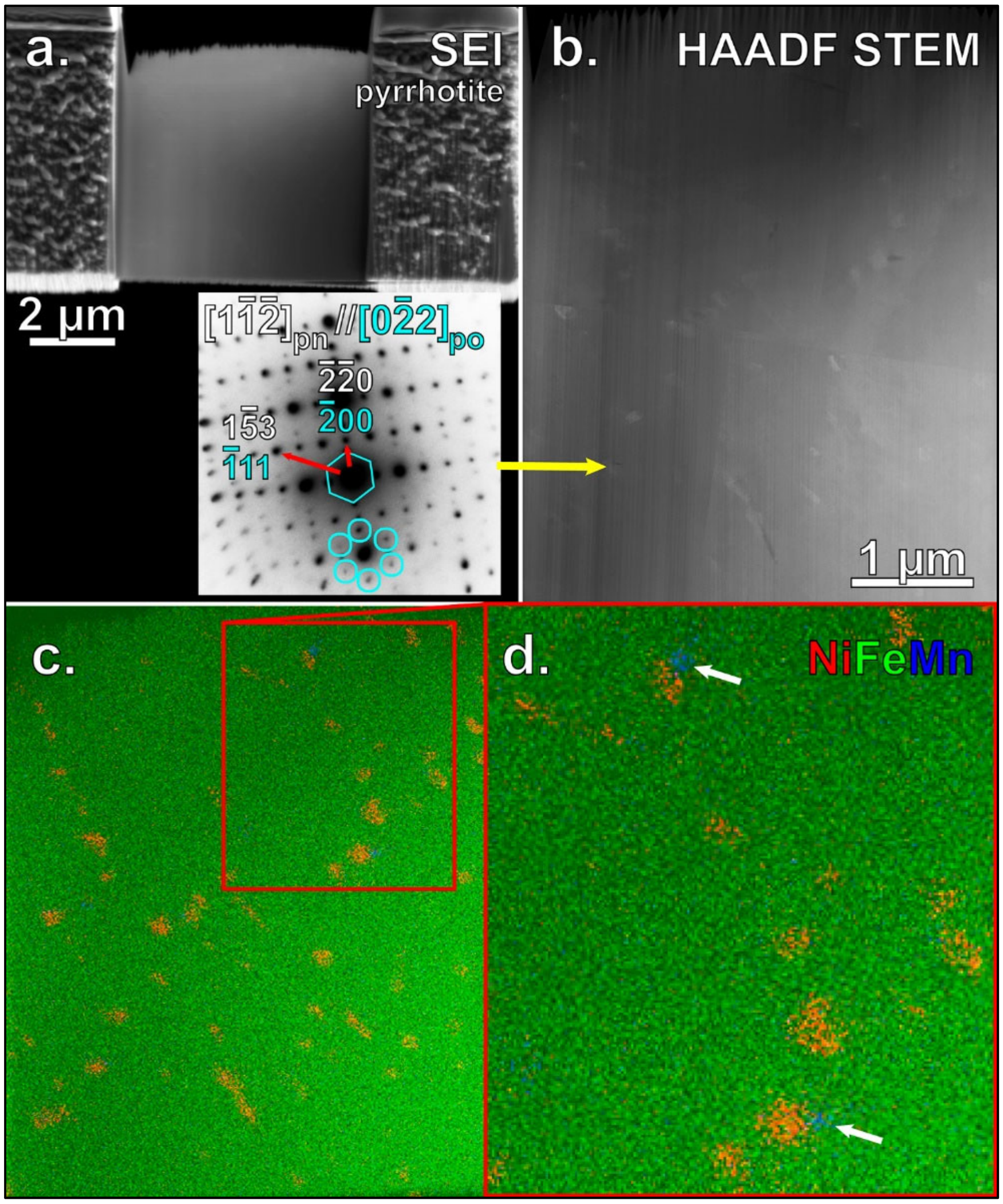

**Figure 4.** SEM and TEM images of pyrrhotite from a SRM grain located in a type IA chondrule of CR EET 92042. BSE images showing the textural context of the grain and the grain itself, along with the location of the FIB section extraction, are presented in Figure 2e and 2f, respectively. An SEI (a) shows the overall FIB section with a small region in the center thinned to electron transparency. A HAADF STEM image (b) shows a higher magnification image of the thinned portion with pentlandite visible as small, slightly brighter inclusions. An electron diffraction pattern is also included from a pentlandite inclusion and shows the presence of the pentlandite (bright diffraction spots) in addition to the adjacent pyrrhotite (faint diffraction spots emphasized with teal circles/hexagon). Composite STEM EDS X-ray maps of Ni (red), Fe (green), and Mn (blue) showing (c) the same field of view as (b) and a higher magnification image (d) of a region with Mn-bearing inclusions (emphazised by the white arrows).

### 3.2. Trace Element Compositions

The SXRF microprobe data of individual spot analyses are summarized in Table 2. All data are in ppm unless otherwise noted. Iron from EPMA analyses was used for standardization of the SXRF spectra, and the Ni abundance from EPMA analyses was used as an additional check and to refine the standardization for samples with high Ni, hence the fixed values in Table 2's SXRF-derived Ni abundances for CM-PPI-Pn and CR-PPI-Pn. As mentioned in the Methods, we opted to use EPMA rather than SXRF data for Co abundances. Individual SXRF spot analyses are listed, with their uncertainties in parentheses. Below each set of spot analyses are the weighted averages and standard deviations for each grain; see the Supplemental Materials for a discussion of how these values were calculated. Trace element data for individual spot analyses of FIB sections from each grain are summarized in Figure 5, with logarithmic plots (Figs. 5b and 5d) included to better show data with low abundances (those near the axes in Figs. 5a and 5c).

Copper concentrations (Fig. 5a–b) vary amongst the different phases with metal (790–950 ppm), pyrrhotite (162–1380 ppm), and pentlandite (510–1660 ppm) ranges overlapping. CR-SRM-Po and CR-PPI-Po grains, which have similar Cu concentrations, appear to have lower Cu contents compared to CM-PPI-Po, while the reverse is true for Cu in CM- or CR-PPI-Pn. Selenium, on the other hand, shows distinct ranges depending on the phase; Se content increases from metal (7.3–19.9 ppm) to pyrrhotite (47–240 ppm) to pentlandite (300–1030 ppm). CR-PPI-Po appears to have lower Se contents compared to CM-PPI-Po, while CR-PPI-Pn appears to have higher Se contents compared to CM-PPI-Pn. CR-SRM-Po has higher Se contents than CM- or CR-PPI-Po.

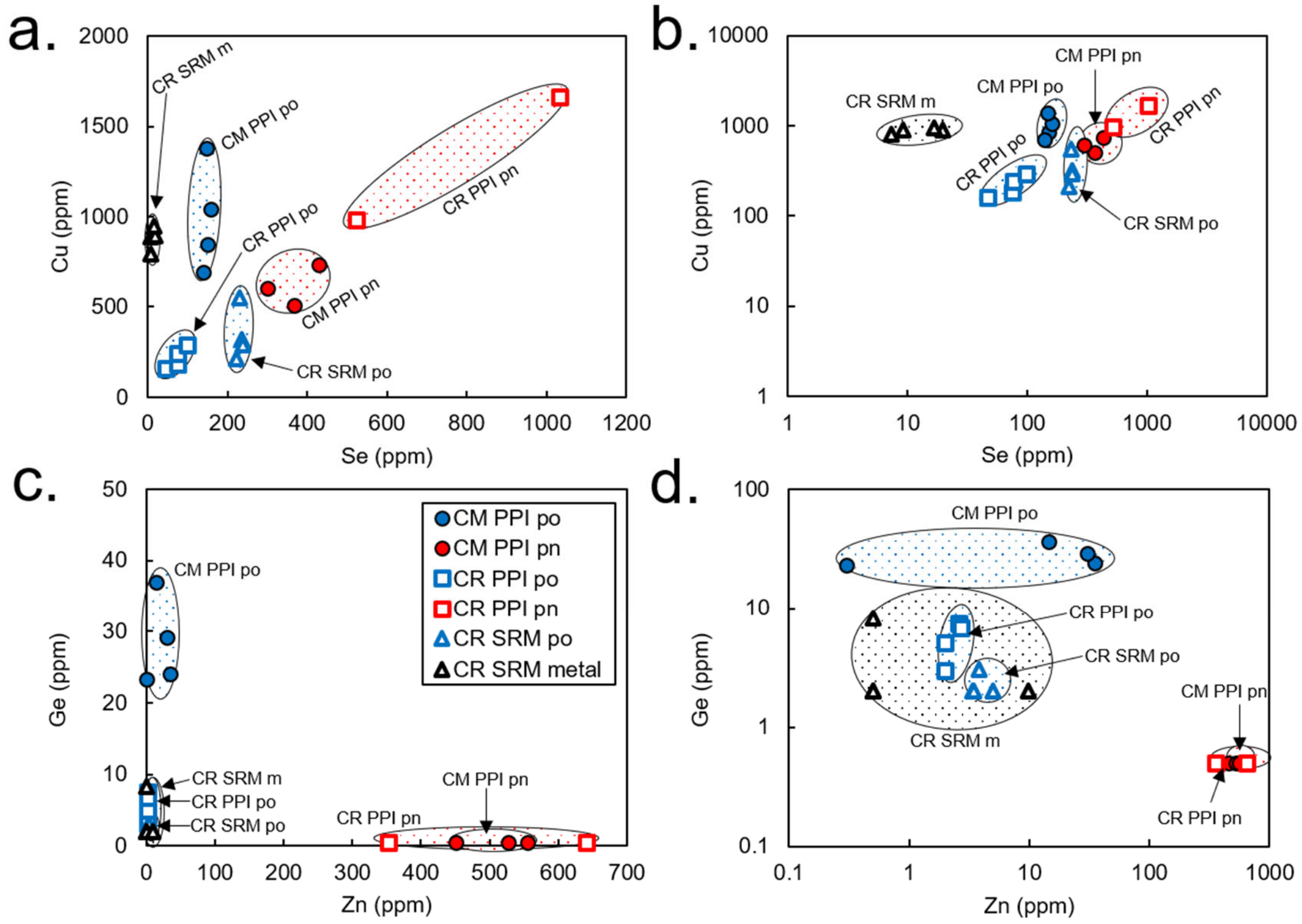

**Figure 5.** Element-element plots (Cu, Se, Zn, and Ge in ppm) of individual spot SXRF analyses of the FIB sections featured in this study, which include sulfides and Fe,Ni metal in CM and CR chondrites. (a–b) show Cu versus Se and (c–d) show Ge versus Zn, with (b) and (d) on logarithmic scales to better show data with low abundances. Distinct groupings of analyses from different FIB sections are apparent in (a–b) as are the different ranges in Se concentration for the different phases. Selenium content increases from metal to pyrrhotite to pentlandite. The different affinities for Ge and Zn in pyrrhotite and pentlandite are apparent in (c–d). Comparatively, pyrrhotite has lower Zn and higher Ge concentrations than pentlandite. Symbols: CM = closed, CR = open, SRM = triangles, CM PPI = circles, CR PPI = squares, Fe,Ni metal = black, pyrrhotite = blue, pentlandite = red.

**Table 2.** Elemental abundances (in ppm unless otherwise noted) for the six FIB sections studied. Errors for each value are presented within parentheses.

| Sample | Phase | Spectrum | Fe[1,2] | S[1, 2] | Ni[2] | Co[2] | Cu | Ge | Zn | Se | Total[1] |
|---|---|---|---:|---:|---:|---:|---:|---:|---:|---:|---:|
| QUE97990 (CM2, PPI, C12S2) | Po | Spot1 | 61.80 (0.18) | 36.77 (0.14) | 12,500 (1,600) | 1,900 (200) | 850 (42) | 24.0 (1.2) | 34.9 (1.8) | 150 (8) | 100.3 (0.3) |
| | | Spot2 | 61.80 (0.18) | 36.77 (0.14) | 12,500 (1,600) | 1,900 (200) | 1,040 (52) | 23.3 (1.2) | <0.3 | 159 (8) | 100.2 (0.3) |
| | | Spot3 | 61.80 (0.18) | 36.77 (0.14) | 12,400 (1,700) | 1,900 (200) | 690 (35) | 37.0 (1.9) | 14.5 (1.1) | 140 (7) | 100.2 (0.3) |
| | | Spot4 | 61.80 (0.18) | 36.77 (0.14) | 12,400 (1,700) | 1,900 (200) | 1,380 (69) | 29.1 (1.7) | 30.5 (2.2) | 148 (8) | 100.2 (0.3) |
| | | Average[3] | 61.80 (0.18) | 36.77 (0.14) | 12,500 (1,600) | 1,900 (200) | 870 (23) | 26.5 (0.7) | 20.0 (0.2) | 149 (4) | 100.2 (0.3) |
| | | Std Dev[3] | N/A | N/A | 58 | N/A | 300 | 6.3 | 16 | 7.8 | |
| | Pn | Spot2 | 40.42 (0.07) | 33.38 (0.25) | 247,000 (910) | 7,000 (300) | 510 (160) | <0.5 | 450 (32) | 370 (30) | 99.4 (0.3) |
| | | Spot3 | 40.42 (0.07) | 33.38 (0.25) | 247,000 (910) | 7,000 (300) | 740 (190) | <0.5 | 530 (37) | 430 (35) | 99.4 (0.3) |
| | | Spot4 | 40.42 (0.07) | 33.38 (0.25) | 247,000 (910) | 7,000 (300) | 610 (190) | <0.5 | 560 (39) | 300 (37) | 99.4 (0.3) |
| | | Average[3] | 40.42 (0.07) | 33.38 (0.25) | 247,000 (910) | 7,000 (300) | 610 (100) | 0.3 (0.1) | 500 (21) | 360 (17) | 99.4 (0.3) |
| | | Std Dev[3] | N/A | N/A | N/A | N/A | 110 | N/A | 55 | 65 | |
| QUE99177 (CR2, PPI, C1S2) | Po | Spot1 | 62.49 (2.31) | 36.91 (0.17) | 8,680 (100) | 1,400 (100) | 162 (8) | <3 | 2.0 (0.4) | 47.0 (2.4) | 100.4 (2.3) |
| | | Spot2 | 62.49 (2.31) | 36.91 (0.17) | 7,980 (100) | 1,400 (100) | 182 (9) | 7.5 (0.5) | 2.6 (0.5) | 74.6 (3.8) | 100.4 (2.3) |
| | | Spot3 | 62.49 (2.31) | 36.91 (0.17) | 6,700 (100) | 1,400 (100) | 240 (12) | 5.1 (0.3) | 2.0 (0.5) | 75.9 (3.9) | 100.2 (2.3) |
| | | Spot4 | 62.49 (2.31) | 36.91 (0.17) | 7,580 (100) | 1,400 (100) | 290 (15) | 6.9 (0.5) | 2.7 (0.6) | 98.9 (5.1) | 100.3 (2.3) |
| | | Average[3] | 62.49 (2.31) | 36.91 (0.17) | 7,740 (100) | 1,400 (100) | 198 (5) | 5.2 (0.1) | 2.2 (0.2) | 64.0 (2.0) | 100.3 (2.3) |
| | | Std Dev[3] | N/A | N/A | 830 | N/A | 58 | 2.3 | 0.4 | 21 | |
| | Pn | Spot3 | 39.64 (0.97) | 33.76 (0.84) | 242,100 (6,600) | 8,000 (200) | 1,660 (330) | <0.5 | 640 (55) | 1,030 (82) | 98.8 (1.4) |
| | | Spot4 | 39.64 (0.97) | 33.76 (0.84) | 242,100 (6,600) | 8,000 (200) | 980 (160) | <0.5 | 350 (28) | 520 (33) | 98.6 (1.4) |
| | | Average[3] | 39.64 (0.97) | 33.76 (0.84) | 242,100 (6,600) | 8,000 (200) | 1,110 (140) | 0.3 (0.1) | 410 (25) | 590 (30) | 98.7 (1.4) |
| | | Std Dev[3] | N/A | N/A | N/A | N/A | 480 | N/A | 200 | 360 | |

| Sample | Phase | Spectrum | Fe[1, 2] | S[1, 2] | Ni[2] | Co[2] | Cu | Ge | Zn | Se | Total[1] |
|---|---|---|---|---|---|---|---|---|---|---|---|
| EET92042 (CR2, SRM, C3S3) | Po | Spot1 | 62.74 (0.15) | 35.97 (0.14) | 7,600 (780) | 800 (200) | 210 (11) | <2 | 5.0 (0.5) | 220 (11) | 99.7 (0.2) |
| | | Spot2 | 62.74 (0.15) | 35.97 (0.14) | 8,840 (780) | 800 (200) | 550 (28) | <2 | 3.4 (0.5) | 230 (12) | 99.8 (0.2) |
| | | Spot3 | 62.74 (0.15) | 35.97 (0.14) | 7,540 (780) | 800 (200) | 320 (16) | <2 | 3.4 (0.5) | 240 (12) | 99.7 (0.2) |
| | | Spot4 | 62.74 (0.15) | 35.97 (0.14) | 6,710 (780) | 800 (200) | 290 (15) | 3.1 (0.4) | 3.8 (0.5) | 240 (12) | 99.6 (0.2) |
| | | Average[3] | 62.74 (0.15) | 35.97 (0.14) | 7,670 (780) | 800 (200) | 280 (7) | 1.5 (0.2) | 3.9 (0.2) | 230 (6) | 99.7 (0.2) |
| | | Std Dev[3] | N/A | N/A | 880 | N/A | 150 | 1 | 0.7 | 7.5 | |
| | M | Spot1 | 92.63 (0.22) | nd | 63,200 (1,000) | 3,100 (100) | 890 (450) | 8.3 (0.5) | <0.5 | 9.1 (0.6) | 101.0 (0.2) |
| | | Spot2 | 92.63 (0.22) | nd | 63,700 (1,000) | 3,100 (100) | 790 (400) | <2 | <0.5 | 7.3 (0.6) | 99.9 (0.2) |
| | | Spot3 | 92.63 (0.22) | nd | 64,800 (1,000) | 3,100 (100) | 890 (450) | <2 | <0.5 | 19.9 (1.3) | 99.9 (0.2) |
| | | Spot4 | 92.63 (0.22) | nd | 64,900 (1,000) | 3,100 (100) | 950 (470) | <2 | 9.9 (1.1) | 16.6 (1.6) | 99.8 (0.2) |
| | | Average[3] | 92.63 (0.22) | nd | 64,100 (1,000) | 3,100 (100) | 870 (22) | 2.8 (0.2) | 2.7 (0.1) | 9.8 (0.4) | 100.2 (0.2) |
| | | Std Dev[3] | N/A | N/A | 840 | N/A | 65 | 3.6 | 4.9 | 6 | |

PPI = pyrrhotite-pentlandite intergrowth, SRM = sulfide-rimmed metal, po= pyrrhotite, pn = pentlandite, m =metal, N/A = not applicable, nd = not detected.

[1]Data are in weight percent element

[2]Elemental abundances were wholly (Fe, S, and Co) or partly (Ni) determined with EPMA. SXRF standardization used Fe for all samples in addition to Ni for pentlandite, hence the fixed values.

[3]The average and standard deviation (Std Dev) are weighted and calculated from unrounded data. See Supplemental Materials for more detail.

Germanium and Zn concentrations (Fig. 5c–d) differ markedly depending on the phase. Comparatively, pyrrhotite (<0.3–34.9 ppm Zn and <2–37 ppm Ge) has lower Zn and higher Ge concentrations than pentlandite (350–640 ppm Zn and <0.5 ppm Ge). Metal has low concentrations of both elements (<0.5–9.9 ppm Zn and <2–8.3 ppm Ge). CM-PPI-Po has higher Ge contents than CR-PPI-Po or CR-SRM-Po, but the PPI-Po Zn contents overlap between the two meteorite groups. The distinct behavior of these two elements between pyrrhotite and pentlandite provides a method of identifying which SXRF spot analyses may have overlapped with submicron inclusions; that is, if a pyrrhotite analysis contains a higher concentration of Zn and a lower concentration of Ge, it likely contains pentlandite inclusions, whereas the reverse is true for pentlandite that contains pyrrhotite inclusions. As Figures 3c, 3f, and 4c–d show, inclusions of one phase within the other do occur. Several such spot analyses show both types of behavior for Zn and Ge compared to other spot analyses of the same grain. These analyses were excluded from any tables, figures, and further discussion. We assume that the remaining analyses lack significant inclusion contribution

Figure 5 shows a few noteworthy differences between the same phases of different textural and meteorite groups. CR-SRM-Po and CR-PPI-Po have similar concentrations of Cu, Zn, and Ge, but form two distinct groups in Se content (Fig. 5a–d), with the CR-SRM-Po having a higher Se concentration. More striking are the differences between CM and CR PPI sulfides. In Figures 5a–b, CR-PPI-Po and CR-PPI-Pn fall along a linear trend with a positive correlation between Cu and Se, but the same is not true of CM-PPI-Po and CM-PPI-Pn, which each appear to fall along their own trends. This difference between the two PPI grains suggests a distinctive process has occurred in one of these groups, which resulted in different behavior of these trace elements in pyrrhotite and pentlandite.

To better compare concentrations of multiple elements between phases, textural groups, and meteorite groups, we plot the averaged data for each grain, normalized to CI chondrite abundances (using data from Palme et al., 2014), as a function of increasing volatility (50% condensation temperature from Lodders, 2003) and present these in Figures 6–7. Data used for calculating CI-normalized data are included in Table A5. Figure 6 illustrates the overall trends in the behavior of the elements, Figure 7a–b groups like phases together and compares the concentrations between meteorite group (CM v. CR) and textural group (PPI v. SRM), and Figure 7c–e groups like textural groups together and compares the concentrations between coexisting phases (PPI po v. pn, SRM po v. metal).

Figure 6 shows that both the least volatile (Ni, Co, Cu) and the most volatile (Se) elements detected are more abundant and greater than CI chondrite abundances in most cases. Comparing trace element abundances of the same phase in different meteorite and/or textural groups has the potential to shed light on formation mechanisms and whether they are similar or different between these groups. The three pyrrhotite grains (Fig. 7a), irrespective of their textural occurrence (PPI or SRM) or chondrite group (CM and CR), have remarkably similar abundance patterns. The only difference is that the CM-PPI-Po has higher concentrations of nearly all elements. Pentlandite in PPI grains from both the CM and CR samples (Fig. 7b) also have very similar concentrations of the elements, with even less variation than was seen in the pyrrhotite. All elements except Cu and Se essentially overlap, within error.

A comparison of coexisting pyrrhotite and pentlandite trace element abundances within a single grain or assemblage has the potential to shed light on partitioning behavior and potential genetic relationships. PPI grain pyrrhotite and pentlandite have distinctly different abundance patterns; pyrrhotite has higher concentrations of Ge, and pentlandite has higher concentrations of Co, Zn, and Se. These relationships are true of both the CM PPI grain (Fig. 7c) and the CR PPI grain (Fig. 7d). Copper, however, has different concentrations in the two phases between the meteorite groups. In the CM PPI grain, pyrrhotite has a higher concentration of Cu, whereas in the CR PPI grains, pentlandite has a higher concentration. The linear trends defined by the CR, but not the CM, PPI pyrrhotite and pentlandite analyses in Figure 5a–b are related to this difference in partitioning. Relative to CI chondrite, Cu and Se are enriched in both phases, Germanium is depleted in both phases, and Zn is enriched in pentlandite and depleted in pyrrhotite. Pyrrhotite and Fe,Ni metal in the SRM grain (Fig. 7e) have very similar concentration patterns, but the metal has higher concentrations of all elements except Zn and Se.

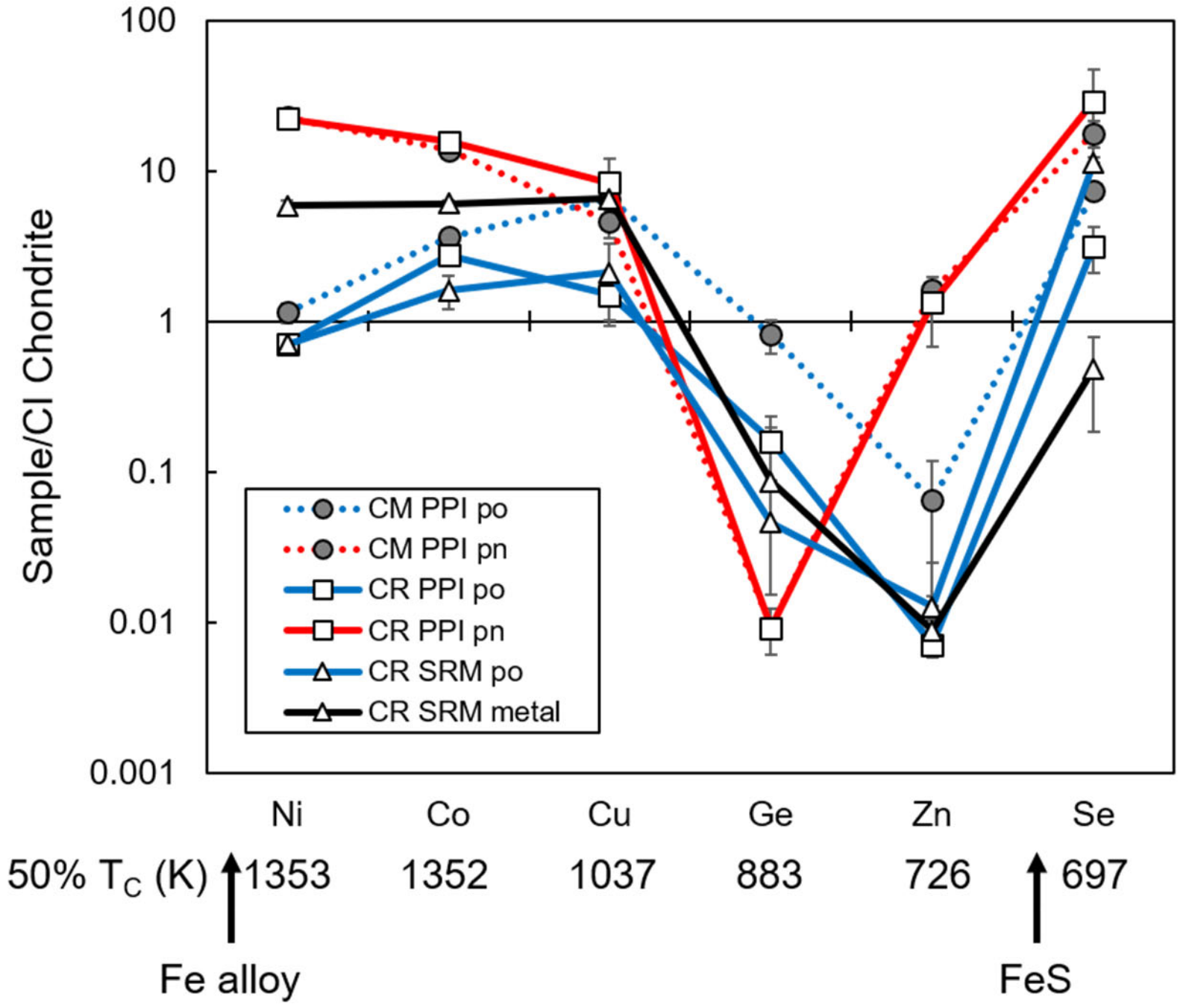

**Figure 6.** Abundances of Ni, minor, and trace elements, normalized to CI chondrite, from SXRF analyses of FIB sections featured in this study, which include sulfides and Fe,Ni metal in CM and CR chondrites. Each line represents an average of multiple spot analyses with error bars, some of which are smaller than the size of the symbol. The line style corresponds to the meteorite group (CM = dashed, CR = solid), the symbol to the textural group (CM PPI = gray circle, CR PPI = white square, and SRM = white triangle), and the line color to the phase (Fe,Ni metal = black, pyrrhotite = blue, pentlandite = red). The 50% condensation temperatures ($T_C$) of the elements are included, as well as the condensation temperatures of Fe alloy and troilite (in K; from Table 8 in Lodders, 2003). Note the enrichments in Ni, Co, Cu, and Se and depletions in Ge and Zn relative to CI chondrite (exceptions include Ni in the CR-PPI-Po and CR-SRM-Po grains and Se in CR-SRM-M grain). The elements of intermediate volatility (Ge and Zn) are depleted relative to CI chondrite values in most cases (exceptions include Zn in CM- and CR-PPI-Pn).

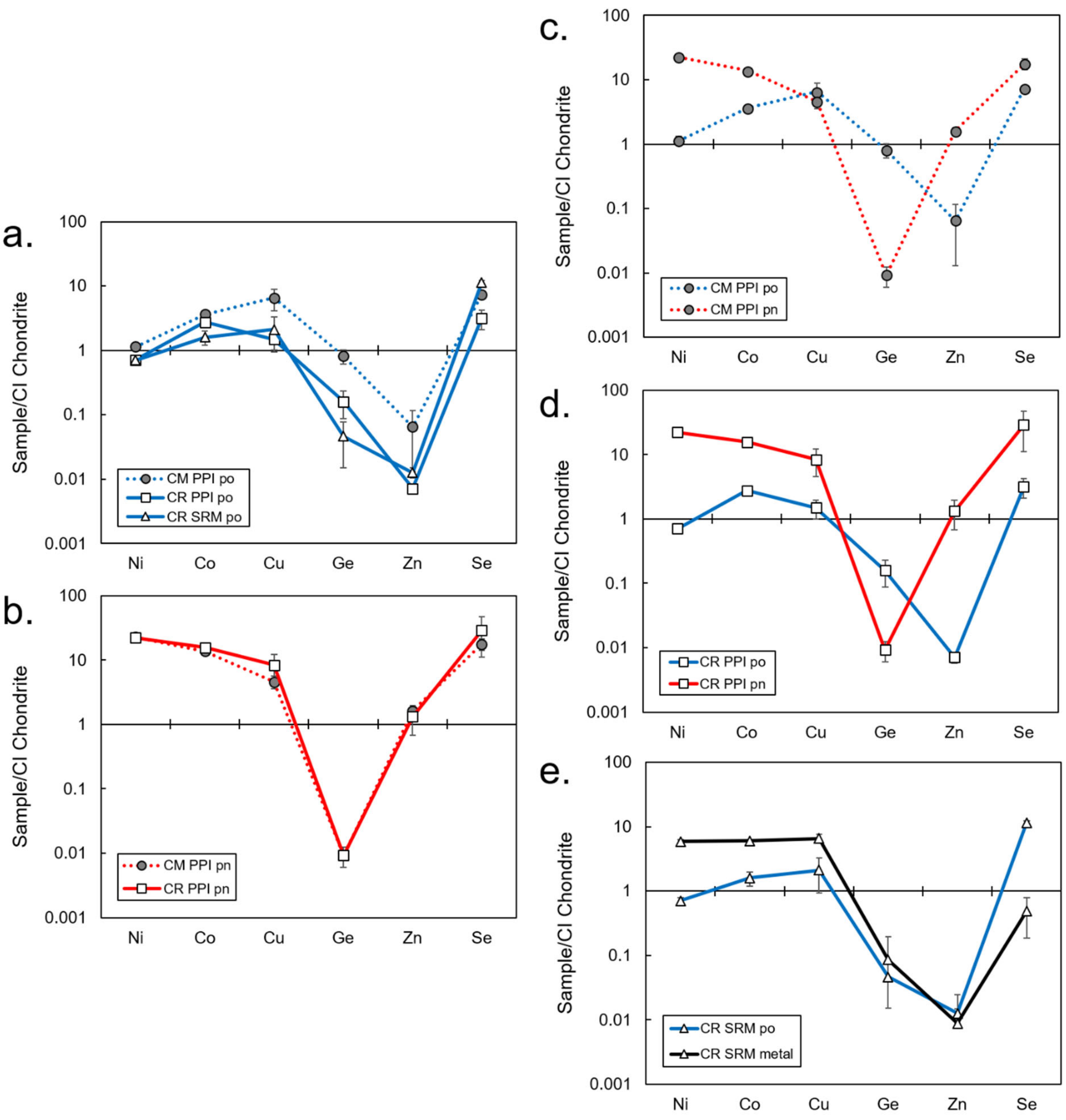

**Figure 7.** CI-normalized abundances as described in Figure 6. (a) compares po data from the CM and CR samples (PPI and SRM po); (b) compares CM and CR PPI pn; (c) compares SXRF data for coexisting po and pn in the PPI grain measured in the CM chondrite; (d) compares coexisting po and pn in the CR PPI grain; and (e) compares coexisting po and metal in the CR chondrite SRM grain. Note the similarities in the abundances between CM and CR PPI phases and in the abundance patterns of the SRM po and metal. All elements, except Cu, behave similarly in (c) and (d) in terms of preference for po or pn. The line form corresponds to the meteorite group (CM = dashed, CR = solid), the symbol to the textural group (CM PPI = gray circle, CR PPI = white square, and SRM = white triangle), and the line color to the phase (Fe,Ni metal = black, pyrrhotite = blue, pentlandite = red).

To summarize our observations from the CI normalized abundance plots (Figs. 6–7), significant observations include:

1) There are generally enrichments relative to CI chondrite in Co, Cu, and Se (Fig. 6);
2) There are generally depletions relative to CI chondrite in Ge and Zn (Fig. 6);
3) CM and CR PPI sulfides (both po and pn) have remarkably similar patterns (Fig. 7a–b), in terms of both their shape and concentration of the elements;
4) Cobalt, Zn, and Se are consistently more abundant in PPI pentlandite, while Ge is more abundant in PPI pyrrhotite (Fig. 7c–d);
5) SRM pyrrhotite and metal have remarkably similar patterns, (Fig. 7e), in terms of both their shape and concentration of the elements.

# 4. DISCUSSION

In this study, we utilized a coordinated analytical approach, the first of its kind, which enabled us to obtain phase-specific trace element concentrations for sulfides and metal coupled with chemical and microstructural data. We have measured the trace element concentrations of six FIB sections in three different iron sulfide grains/sulfide-metal assemblages. These act as representative samples of two of the populations of primary sulfides in CM and CR chondrites discussed in Singerling and Brearley (2018), the PPI grains and the SRM grains. The PPI grains have been proposed to have formed by crystallization of monosulfide solid solution (*mss)* during chondrule formation followed by solid-state unmixing (Singerling and Brearley, 2018 and references therein). In such a process, the PPI grain compositions that we have measured are the result of solid-state elemental partitioning between two different sulfide phases (pyrrhotite and pentlandite) during cooling and do not represent their primary compositions. However, the bulk composition of the *mss* from which they exsolved was established in the solar nebula. The SRM grains, on the other hand, have been proposed to have formed by sulfidization of Fe,Ni metal by $H_2S$ gas in the solar nebula (Singerling and Brearley, 2018).

In the following discussion, we divide the PPI and SRM grains into separate sections and then discuss the previously proposed formation mechanisms for the primary sulfides. This is followed by a summary of the general characteristics of the abundance patterns and the behavior of each of the analyzed elements. We then examine which processes (i.e., condensation, evaporation, melting and crystallization, solid-state partitioning, and/or sulfidization) likely contributed to these patterns and evaluate the previously proposed formation mechanisms for the primary sulfides in the context of the processes indicated by trace element concentrations. Figure 8 summarizes our discussion of the processes and their effects on trace elemental abundances for both the PPI (Fig. 8a) and SRM (Fig. 8b) grains.

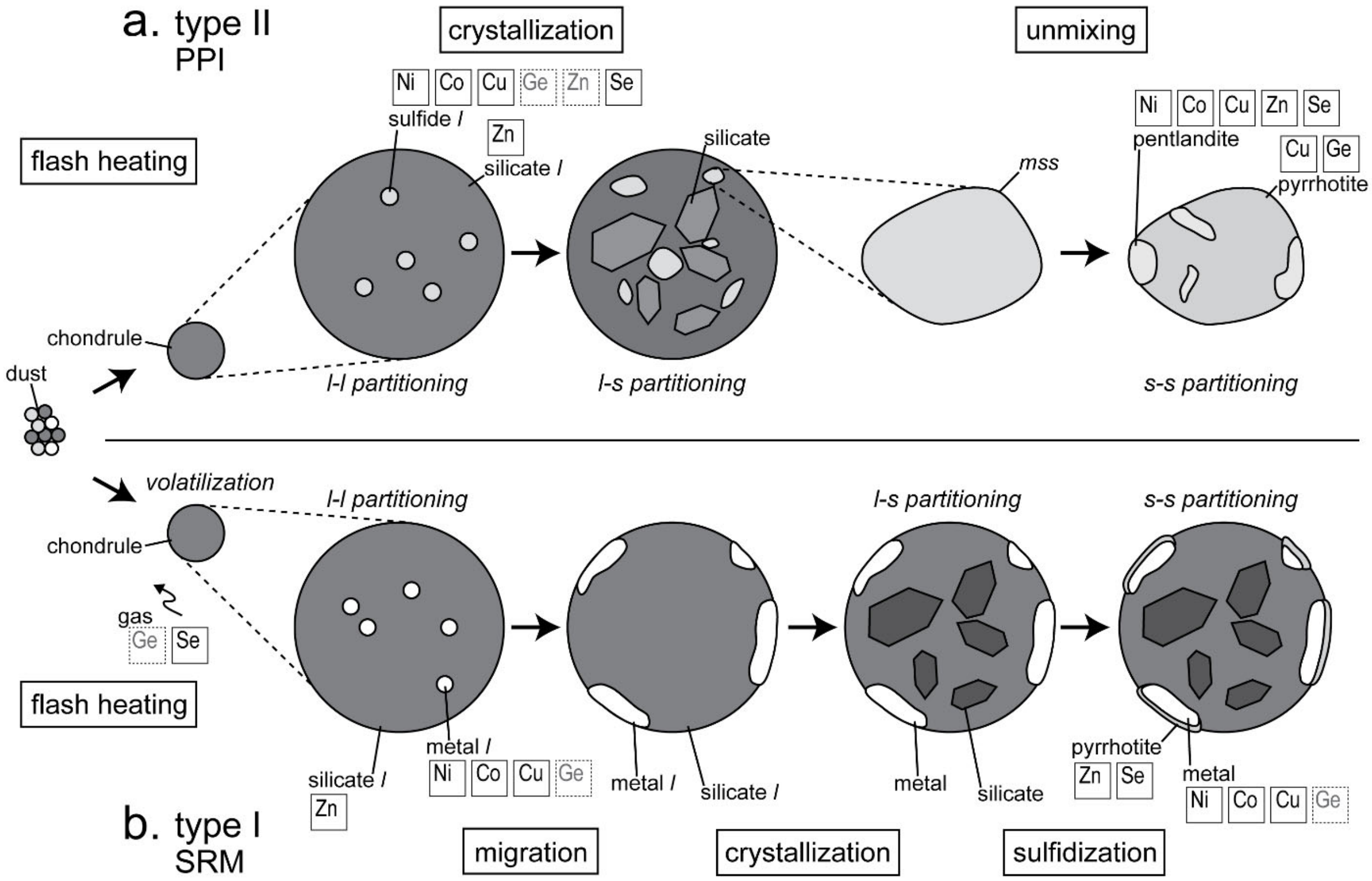

**Figure 8.** Summary of the processes and their effects on partitioning of the trace elements for (a) type II chondrules, which applies to the PPI grains, and (b) type I chondrules, which applies to the SRM grains. Significant processes include (a) liquid-liquid partitioning between sulfide and silicate melts and solid-state unmixing of pyrrhotite and pentlandite from *mss* for the PPI grains, whereas (b) liquid-liquid partitioning between metal and silicate melts and sulfidization of metal are most pertinent for the SRM grains. The elements are listed next to the phase they most strongly partition into; elements listed for both phases partition into both. The partitioning behavior of elements with dashed lines and gray text are less certain. See text for additional discussion.

## 4.1. PPI Grain Trace Element Patterns and Abundances

The PPI grains, in both the CM and CR chondrites, are postulated to have formed in the solar nebula (e.g., Harries and Langenhorst, 2013; Schrader et al., 2015; 2016; Singerling and Brearley, 2018). The following is a brief summary of the processes involved in their formation that we refer to as the crystallization model (see Singerling and Brearley, 2018 for more detail):

1) Aggregation of dust consisting of silicate, metal, and sulfide precursors surrounded by gases in the solar nebula.
2) Formation of chondrules by flash heating with retention of most of the volatile elements.
3) Formation of immiscible silicate and sulfide melts within type IIA chondrules.
4) Crystallization of *mss* (commencing at ~1100°C; Kullerud, 1963), as well as crystallization of silicate phases.
5) Solid-state unmixing of *mss* into pyrrhotite and pentlandite at ~800°C (Kitakaze et al., 2011).

6) Depending on the cooling rate of the system, PPI grains experience different degrees of phase separation. Slower cooling rates allow for greater phase separation and result in greater degrees of solid-state partitioning between pyrrhotite and pentlandite.

As is apparent above, the PPI grains do not represent a primary composition indicative of the solar nebula; however, the composition of the *mss* from which they exsolved does, because the *mss* formed by crystallization of an immiscible melt within chondrules (Singerling and Brearley, 2018). If the proportions of the two phases (i.e., pyrrhotite and pentlandite) are known, we can use modal recombination analysis (e.g., Berlin et al., 2009; Singerling and Brearley, 2018) to determine the bulk trace element concentrations of the PPI grains, that is, the trace element concentrations of the original mss (Table A4). The area fraction of pyrrhotite and pentlandite portions of the PPI grains, as determined using the thresholding function on BSE images within ImageJ, are 70±4% pyrrhotite and 30±2% pentlandite for the CM PPI grain and 46±5% pyrrhotite and 54±5% pentlandite for CR PPI grain. Figure 9 illustrates the bulk composition of the PPI grains, equivalent to the *mss*, along with the range as defined by the pyrrhotite and pentlandite trace element concentrations. The ranges shown in Figure 9 include the two endmembers, pyrrhotite and pentlandite, for this calculation.

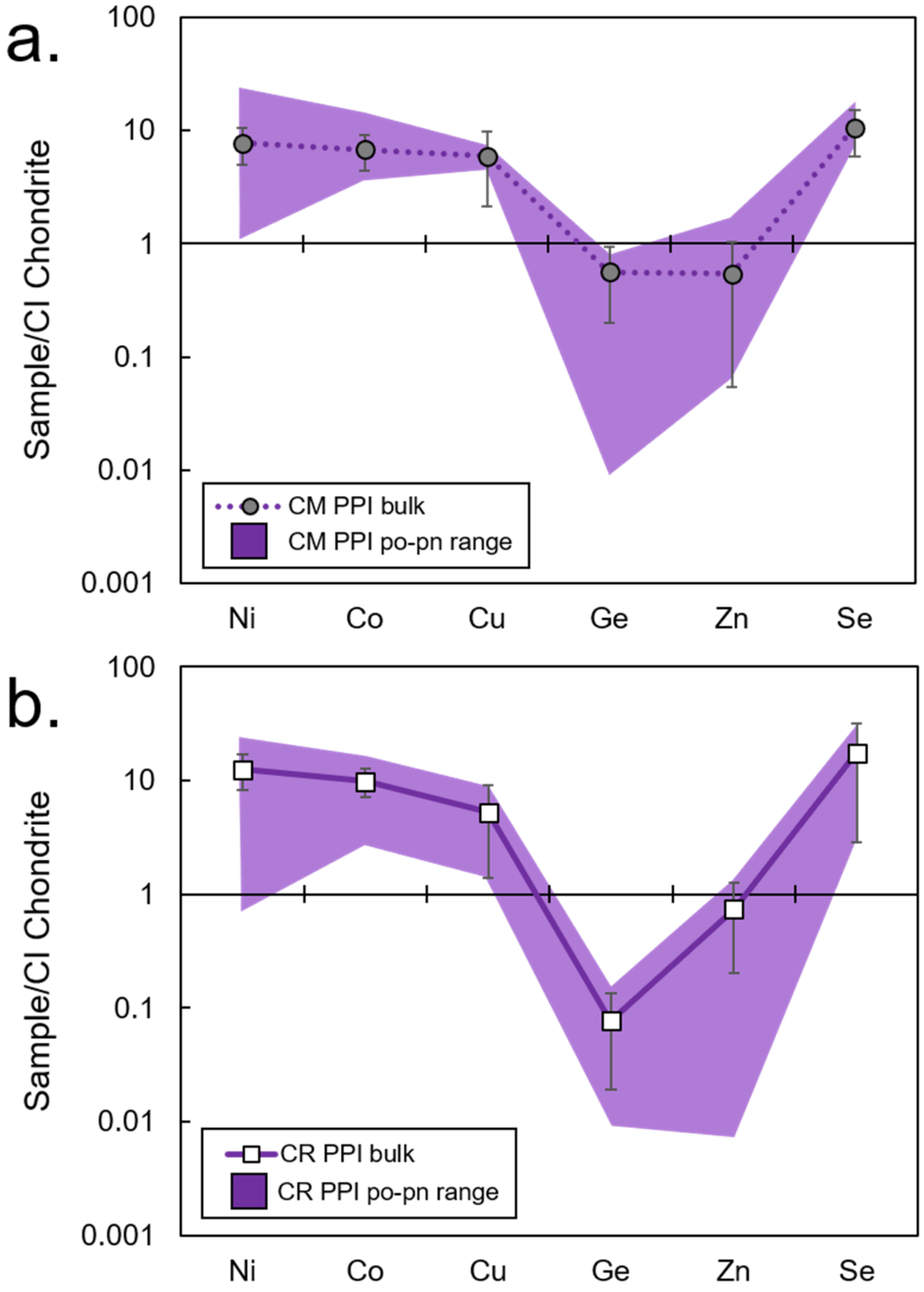

**Figure 9.** CI-normalized abundance plot, as described in Figure 6, showing a modal recombination of SXRF data for the bulk compositions of the (a) CM PPI grain and (b) CR PPI grain. Both plots also include the range of concentrations defined by pyrrhotite and pentlandite from each grain as purple envelopes. Bulk refers to the bulk composition of the PPI, which we assume is equivalent to the composition of the *mss*.

The enrichments in Co, Cu, and Se in the *mss* are consistent with the chalcophile nature of these elements. Cobalt and Cu can substitute for Fe, while Se can substitute for S in sulfide minerals due to their similar ionic radii and valences (Cook et al., 2009). The patterns for Ge and Zn are more complicated, however. The concentration of Ge in the *mss* is slightly depleted relative to CI chondrite; it is depleted both in pyrrhotite and pentlandite, though it is only slightly depleted in the former and strongly depleted (i.e., below detection limit) in the latter. The concentration of Zn is comparable to Ge for the *mss*; that is, it is slightly depleted relative to CI. However, contrary to Ge, Zn shows different behavior between pyrrhotite and pentlandite. It is depleted in pyrrhotite and enriched in pentlandite. Overall, the pyrrhotite, pentlandite, and *mss* patterns of the CM and CR PPI grains are remarkably similar (e.g., Figs. 7 and 9) implying that the grains from two different meteorite groups share similar formation mechanisms. We next determine if the proposed processes responsible for the formation of the PPI grains can explain the observed complexities in the trace element concentrations.

Significant processes which likely operated on and were responsible for the formation of the PPI grains include: volatilization during chondrule formation, formation of immiscible silicate and *mss* melts (liquid-liquid partitioning), crystallization of silicate solids and *mss* (liquid-solid partitioning), and unmixing into pyrrhotite and pentlandite (solid-solid partitioning). When discussing such processes and the expected affinities of the elements for different phases, the elements' behaviors can be broadly divided into two categories—cosmochemical and geochemical. The cosmochemical behavior of an element is a function of the condensation temperature of the host phases the element is incorporated into, whereas the geochemical behavior is a function of the affinities of the elements to form certain phases—lithophile tending to form silicates or oxides, siderophile tending to form metal alloys, and chalcophile tending to form sulfides. Cosmochemical behavior of the elements is significant for volatilization during chondrule formation; however, geochemical behavior dominates for the other processes mentioned above (i.e., immiscibility, crystallization, unmixing). Table 3 summarizes these different behaviors and the parameters responsible for them.

We now discuss how these different processes could have affected the concentrations of the elements in the different stages of formation of the PPI grains. Chondrule formation could have potentially removed some proportion of the more volatile elements (Ge, Zn, Se) from chondrules. However, similar to the case with Na and S, short heating times, high initial cooling rates, high dust/gas ratios, and high oxygen fugacity likely preserved some of the volatile elements in type II chondrules (Yu et al., 1996). Whatever the cause, the presence of S and Na in chondrules indicates volatile retention, which would likely extend to other elements such as Ge, Zn, and Se for the PPI grains. It is important to point out that the volatility of these elements may vary based on the gas composition as well as the partial pressures of constituent compounds, with each being affected differently by flash heating events. For example, in interplanetary dust particles (IDPs), Zn has been observed to be depleted even relative to other volatile elements, such as Ge and Se (Fig. 7 in Flynn and Sutton, 1992). This behavior was attributed to atmospheric heating by Flynn and Sutton (1992) and suggests that Zn was more volatile in this process than Ge and Se. The same could be true for flash heating events during chondrule formation.

**Table 3.** Cosmochemical and geochemical behaviors of the elements detected in sulfide and metal SXRF analyses.

| | | Cosmochemical | Geochemical |
|---|---|---|---|
| | 50% $T_C$ (K)[1] | Behavior(s) | Dominant Behavior |
| Fe[2] | 1357 | Siderophile, lithophile, chalcophile | Siderophile |
| Ni | 1353 | Siderophile | Siderophile |
| Co | 1352 | Siderophile | Siderophile |
| Cu | 1037 | Siderophile | Siderophile |
| Ge | 883 | Siderophile | Siderophile |
| Zn | 726 | Lithophile, chalcophile | Chalcophile |
| S[2] | 704 | Chalcophile | Chalcophile |
| Se | 697 | Chalcophile | Chalcophile |

[1]50% $T_C$ = 50% condensation temperature in K at $1\times10^{-4}$ bar.
[2]$T_C$ for Fe and S are for Fe alloy and troilite, respectively.
Sources: Cosmochemical data and behavior from Lodders (2003); geochemical behavior from McSween and Huss (2010).

Immiscible melt formation would have partitioned elements into either silicate or *mss* fractions in the type II chondrules. MacLean and Shimazaki (1976) performed partitioning experiments on iron sulfide and iron silicate melts (1 bar, 1150°C, and low $fO_2$ and $fS_2$; quantitative parameters for the fugacities were not provided) and found that Ni, Co, and Cu partitioned into the sulfide melt, whereas Zn partitioned into the silicate melt. Experimental work on olivine, silicate, and sulfide melts by Gaetani and Grove (1997) mimicked chondrule formation conditions (1 atm, 1350°C, log $fO_2$ from -7.9 to -10.6, log $fS_2$ from -1.0 to -2.5) and found that Ni, Co, and Cu always partitioned into the sulfide melt and did so more strongly under conditions of low $fO_2$ and high $fS_2$. They also found that the partitioning behavior of Cr and Mn depended on the $fO_2$ and $fS_2$ of the system, with low $fO_2$ and high $fS_2$ conditions promoting partitioning into the sulfide melt. Unfortunately, experimental work involving the partitioning behavior of Ge between silicate and sulfide melts is not available, so the expected geochemical behavior for Ge (siderophile) in Table 3 is the only guideline at our disposal. For the PPI grains, we would expect *mss* to have enrichments in Ni, Co, and Cu and depletions in Zn. Germanium concentrations are less certain given the lack of experimental data for it. In some cases (low $fO_2$ and high $fS_2$), we might also expect to have Cr and Mn present in *mss* though, with cooling, these are likely to have formed exsolution products, consistent with our observations of Cr- and Mn-bearing inclusions in our TEM studies.

Crystallization of *mss* would have partitioned elements between solid *mss* and residual sulfide melt in the type II chondrules. Experimental work has observed that Se partitions as an anion ($Se^{2-}$) and substitutes for $S^{2-}$ in *mss* (Helmy et al., 2010), while the partitioning behavior of Ni and Cu depend strongly on the bulk S content of the system and the *mss* (Li et al., 1996); greater S contents correspond to stronger partitioning of Ni and Cu into *mss*. This dependence on the bulk S content may explain the variability in the behavior of Cu observed between the CM and CR PPI *mss* (Figs. 9, 12).

Solid-state unmixing of *mss* would have partitioned elements into pyrrhotite and pentlandite. Goldschmidt's rules of substitution are valuable tools in determining the expected partitioning behavior between pyrrhotite and pentlandite, especially given the lack of experimental studies for the trace elements of interest. Table 4 presents the variables useful in

**Table 4.** Data for determining the most likely substitutions of trace elements in pyrrhotite, pentlandite, and kamacite. The percent difference shows the likelihood that other ions/atoms would substitute for those normally present in each structural site. Percent differences greater than 15 (light gray, bolded text) limit substitution.

| **Phase** | **Site** | **Ion/Element** | **CN** | **Crystal/Atomic Radius (Å)[1]** | **%Diff** | |
|---|---|---|---|---|---|---|
| Pyrrhotite | M1 | $Fe^{2+}$ | 6 | 0.78 | **$Fe^{2+}$** | |
| | | $Ni^{2+}$ | 6 | 0.69 | 12 | |
| | | $Co^{2+}$ | 6 | 0.75 | 5 | |
| | | **$Cu^{1+}$** | **4** | **0.60** | **26** | |
| | | $Cu^{2+}$ | 6 | 0.73 | 7 | |
| | | $Zn^{2+}$ | 6 | 0.74 | 5 | |
| | Vacancies | $Fe^{3+}$ | 6 | 0.65 | **$Fe^{3+}$** | |
| | | $Ni^{3+}$ | 6 | 0.56 | 14 | |
| | | **$Co^{3+}$** | **6** | **0.55** | **17** | |
| | | $Cu^{1+}$ | 4 | 0.60 | 8 | |
| | | **$Cu^{3+}$** | **6** | **0.54** | **18** | |
| | | **$Ge^{4+}$** | **4** | **0.53** | **20** | |
| | S | $S^{2-}$ | N/A | 1.84 | **$S^{2-}$** | |
| | | $Se^{2-}$ | N/A | 1.98 | 7 | |
| Pentlandite | M1 | $Fe^{2+}$ | 6 | 0.78 | 12 | |
| | | $Ni^{2+}$ | 6 | 0.69 | | |
| | | | | | **$Fe^{2+}$** | **$Ni^{2+}$** |
| | | $Co^{2+}$ | 6 | 0.75 | 5 | 8 |
| | | **$Cu^{1+}$** | **4** | **0.60** | **26** | 14 |
| | | $Cu^{2+}$ | 6 | 0.73 | 7 | 6 |
| | | $Zn^{2+}$ | 6 | 0.74 | 5 | 7 |
| | M2 | $Ni^{2+}$ | 4 | 0.55 | 14 | |
| | | $Fe^{2+}$ | 4 | 0.63 | | |
| | | | | | **$Fe^{2+}$** | **$Ni^{2+}$** |
| | | $Co^{2+}$ | 4 | 0.58 | 8 | 5 |
| | | $Cu^{+1}$ | 4 | 0.60 | 5 | 9 |
| | | $Cu^{2+}$ | 4 | 0.57 | 10 | 4 |
| | | $Zn^{2+}$ | 4 | 0.60 | 5 | 9 |
| | S | $S^{2-}$ | N/A | 1.84 | **$S^{2-}$** | |
| | | $Se^{2-}$ | N/A | 1.98 | 7 | |
| Kamacite | N/A | $Fe^{0}$ | N/A | 1.56 | **$Fe^{0}$** | |
| | | $Ni^{0}$ | N/A | 1.49 | 5 | |
| | | $Co^{0}$ | N/A | 1.52 | 3 | |
| | | $Cu^{0}$ | N/A | 1.45 | 7 | |
| | | **$Ge^{0}$** | **N/A** | **1.25** | **22** | |

CN = coordination number, %Diff = percent difference between the ion in crystal structure (as listed) and the substituting ion, vacancies = M1 sites with vacancy substitution, N/A = not applicable.

[1]Radii of ions are crystal radii from Shannon (1976), whereas radii of elements (neutrally charged) are atomic radii from Clementi et al. (1963).

determining the most likely substitutions within pyrrhotite and pentlandite. Percent differences greater than 15 preclude substitution according to Goldschmidt's rules.

From the percent differences in Table 4, in addition to our knowledge of the basic structure of pyrrhotite and pentlandite, we can estimate the general behavior of the elements in partitioning between the phases and compare it to the actual partitioning observed in the PPI grains. The structures of pyrrhotite and pentlandite are key in this discussion. Pyrrhotite ($Fe_{1-x}S$) essentially has a single M site, which is normally occupied by $Fe^{2+}$. However, as the table clearly shows, other divalent ions can substitute for $Fe^{2+}$. Additionally, $Fe^{3+}$ can substitute if coupled with vacancies; that is, three $Fe^{2+}$ are substituted by two $Fe^{3+}$ and a vacancy. This allows for the possibility of coupled vacancy substitution in pyrrhotite, which more readily accommodates ions with a valence greater than 2+. Pentlandite ($(Fe,Ni)_9S_8$) has two M sites, one tetrahedral (M2) and one octahedral (M1), which occur in an 8:1 ratio. Either can be occupied by $Fe^{2+}$ or $Ni^{2+}$. In Table 4, we show percent differences for both ions for each M site, since each can accommodate different elements during substitution. We have excluded $Ge^{2+}$ (which has octahedral coordination) from the table, because Ge usually occurs in tetrahedral coordination in sulfides owing to the highly covalent nature of the Ge-S bond (Bernstein, 1985). This implies that $Ge^{4+}$ is the preferred valence state in sulfides, substituting for divalent cations such as $Fe^{2+}$ or $Zn^{2+}$ and requiring vacancy formation to maintain charge balance (see later discussion). This appears to be the preferred substitution reaction, at least for Ge in sphalerite (e.g., Bernstein, 1985; Cook et al., 2015; Sahlstrom et al., 2017). For the sulfides, we have also included $Cu^{1+}$ that is tetrahedrally coordinated in the table, as this appears to be the preferred valence and coordination for Cu in many sulfides (e.g., covellite, CuS, which consists of sheets of tetrahedrally coordinated Cu, trigonally coordinated Cu, and covalent S-S dimers (Makovicky, 2006).

Although Goldschmidt's rules provide a useful basis for ion substitution in crystal structures, it may also be necessary to consider more complex mechanisms in order to fully interpret the trace elemental abundances observed. This is especially pertinent for the current study, given that several of the trace elements of interest are transition metals and, as such, have electron configurations characterized by orbital degeneracy. For example, the Jahn-Teller effect, characterized by geometrical distortions in the crystal structure owing to certain electron configurations, is most pronounced for octahedrally coordinated ions. Since the M1 site in pyrrhotite is octahedrally coordinated, pyrrhotite may have a more pronounced Jahn-Teller effect than pentlandite, which has more tetrahedrally coordinated M sites; the M2 site (tetrahedrally coordinated) occurs 8 times for every M1 site (octahedrally coordinated). Sakkopoulos et al. (1984) argue that distortions from the Jahn-Teller effect are negligible in pyrrhotite. However, the extent of the Jahn-Teller effect in pentlandite is not known, and similar mechanisms dependent on electron configurations in either phase could also play a role. More studies are needed in this area.

Table 5 and Figure 10 summarize our element distribution (${}^{\alpha/\beta}D_i$) data for the PPI grains, which is essentially the effective partition coefficient. Partition coefficients are meant to convey the behavior of materials in different media in a system at equilibrium, but we do not know for certain if the PPI grains formed under equilibrium. However, given that diffusion rates tend to be rapid in sulfides, the PPI grains likely formed under equilibrium conditions, at least down to the temperature at which closure for diffusion occurs. Additionally, there is no evidence for chemical zoning in the sulfides, which would be evidence for disequilbrium. In any case, we make use of the standard notation used for partition coefficients, recognizing that in some cases, this might not represent true equilibrium partitioning. As there is no convention for calculation of

$^{\alpha/\beta}D_i$ in solid phase-solid phase systems for sulfides, we have calculated the values as pentlandite/pyrrhotite (α/β) for the PPI grains. For $D_i$ >1, the element preferentially partitions into pentlandite, and for $D_i$ <1, the element preferentially partitions into pyrrhotite.

**Table 5.** Element distributions ($D_i$) and data used for the calculations for CM and CR PPI pentlandite/pyrrhotite and SRM metal/pyrrhotite. All elemental concentrations are in ppm unless otherwise noted. Errors for each $D_i$ are presented within parentheses; errors for abundances can be found in Table 2.

| Sample | Phase | Ni[1] | Co[1] | Cu | Ge | Zn | Se |
|---|---|---|---|---|---|---|---|
| QUE 97990 (CM2, PPI, C12S2) | Pn | 24.7 | 0.187 | 610 | 0.3 | 500 | 360 |
| | Po | 1.25 | 0.703 | 870 | 26.5 | 20.0 | 149 |
| | $^{pn/po}D_i$ | 19.8 (2.6) | 3.77 (0.43) | 0.70 (0.27) | 0.011 (0.005) | 25.0 (20.2) | 2.42 (0.45) |
| QUE 99177 (CR2, PPI, C1S2) | Pn | 24.2 | 0.804 | 1,110 | 0.3 | 410 | 590 |
| | Po | 0.774 | 0.141 | 198 | 5.2 | 2.2 | 64.0 |
| | $^{pn/po}D_i$ | 31.3 (3.5) | 5.70 (0.43) | 5.61 (2.93) | 0.06 (0.03) | 186 (97) | 9.22 (6.39) |
| EET 92042 (CR2, SRM, C3S3) | M | 6.41 | 0.082 | 870 | 2.8 | 2.7 | 9.8 |
| | Po | 0.767 | 0.309 | 280 | 1.5 | 3.9 | 230 |
| | $^{m/po}D_i$ | 8.36 (0.97) | 3.78 (0.93) | 3.11 (1.68) | 1.87 (2.7) | 0.69 (1.26) | 0.04 (0.03) |

PPI = pyrrhotite-pentlandite intergrowth, SRM = sulfide rimmed metal, $D_i$ = element distributions (calculated from unrounded data), po = pyrrhotite, pn = pentlandite, m = metal.
[1]Data are in weight percent element.

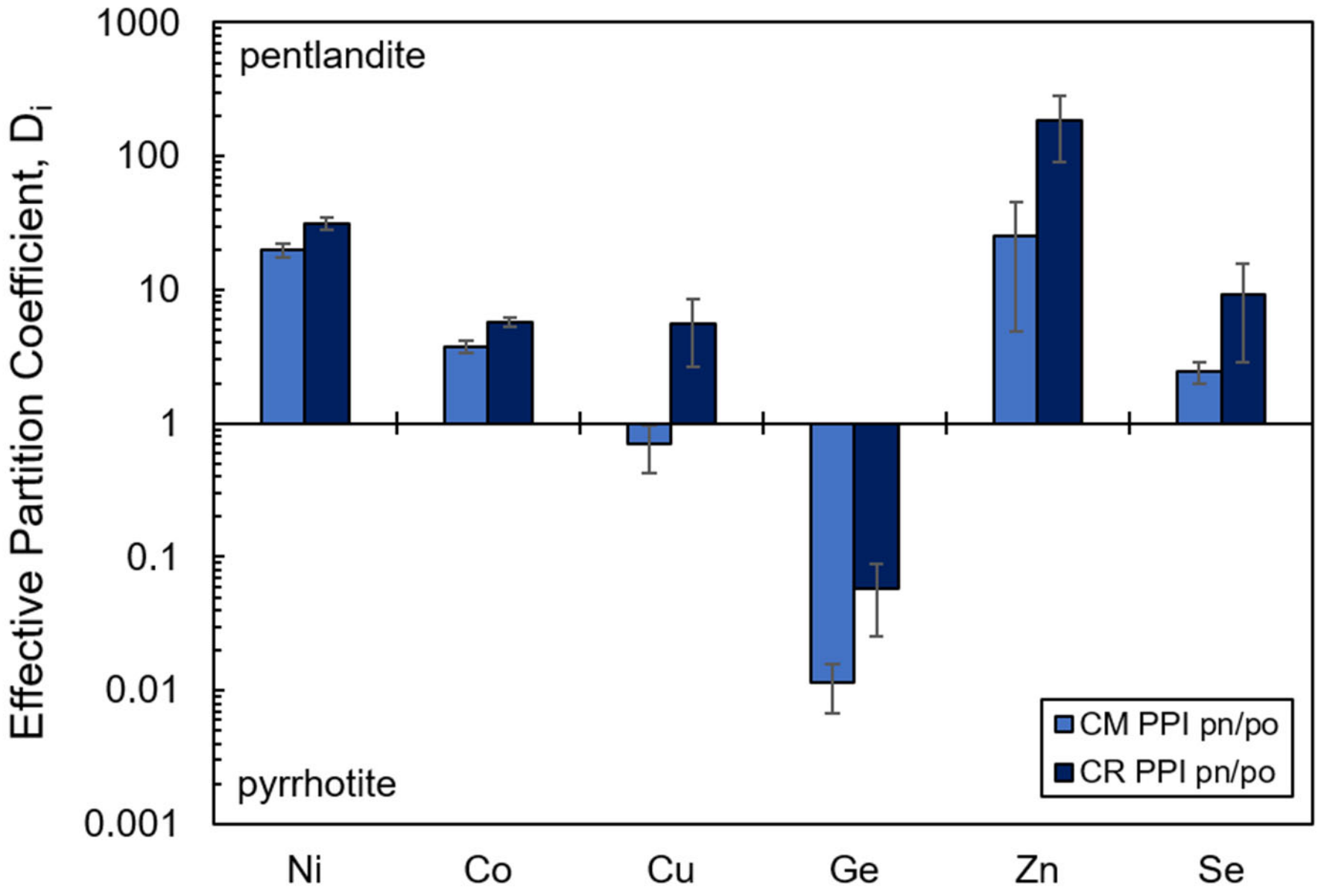

**Figure 10.** Bar graph illustrating the effective partition coefficients for pentlandite v. pyrrhotite in the CM PPI grain and pyrrhotite v. pentlandite in the CR PPI grain. All elements except for Ge, and partially Cu, are preferentially incorporated into pentlandite over pyrrhotite. The deviation of $D_i$ from unity is greater for the CR PPI grain compared to the CM PPI grain for all elements except Ge. This implies that the CR PPI grain has experienced more extensive phase separation likely from a slower cooling rate and, hence, to a lower closure temperature.

With the structural substitution considerations in mind, we can explain some aspects of the pyrrhotite and pentlandite partitioning behaviors in the PPI grains. Pentlandite has greater concentrations of Co compared to pyrrhotite. As $Co^{2+}$ has a similar ionic radius to both $Fe^{2+}$ and $Ni^{2+}$ in the M1 and M2 sites in both minerals (percent differences of 5–8% for all), differences in ionic charge radii is not the only factor affecting substitution. Instead, electronegativity of the elements may play a role. Comparing electronegativites of Co and Ni (1.88 and 1.91, respectively) versus Co and Fe (1.88 and 1.83, respectively), we find that the percent difference is lower between Co and Ni (1.58%) than between Co and Fe (2.70%). Since Co may preferentially substitute more strongly for Ni than Fe, we would expect it to be in higher abundance in pentlandite given that more of its M sites can be occupied by Ni than is the case for pyrrhotite. That being said, these differences in electronegativities are not large, and so this may not be the dominant factor in Co's higher abundance in pentlandite. Other processes dependent on orbital degeneracy may play a role.

Germanium and Zn, more than any other trace elements, show the strongest partitioning between pyrrhotite and pentlandite and have opposite behaviors, with Ge preferring pyrrhotite and Zn pentlandite. Germanium's preference for pyrrhotite can be explained by its proposed coupled vacancy substitution. In sphalerite, the most extensively-studied Ge-bearing sulfide, Ge has been proposed to substitute for $Zn^{2+}$ by the following mechanism: $2Zn^{2+} \rightarrow Ge^{4+} + \square$ (Bernstein, 1985; Cook et al., 2015; Sahlstrom et al., 2017). This substitution is similar to the $Fe^{3+}$ substitution for $Fe^{2+}$ in pyrrhotite ($3Fe^{2+} \rightarrow 2Fe^{3+} + \square$) that is responsible for the various superstructures of pyrrhotite. In pentlandite, however, there is little evidence that coupled vacancy substitution occurs. This could explain why Ge is below the detection limit in the CM and CR PPI grain pentlandite, but present in the pyrrhotite.

Zinc's preference for pentlandite can be explained by its ability to substitute into both M sites in pentlandite and its inability for vacancy substitution in pyrrhotite. Still, this does not quite explain why the partitioning is so distinct for Zn. Zinc is able to substitute for $Fe^{2+}$ in all M sites in pyrrhotite, so the fact that it is able to do the same in pentlandite does not account for the large difference in partitioning. The difference in proportions of metals (i.e., cations in the M sites) to sulfur between pyrrhotite and pentlandite may also play a role. In pyrrhotite, only the M sites containing $Fe^{2+}$ are available for substitution, whereas those containing $Fe^{3+}$ are not. However, in pentlandite, all the M sites essentially contain $Fe^{2+}$ or $Ni^{2+}$ making them available for substitution. In summary, our observations of trace element concentrations in PPI grain pyrrhotite and pentlandite are consistent with solid-state unmixing for most elements.

Copper shows more complicated behavior; in the CR PPI grain, it has a higher concentration in pentlandite, whereas in the CM PPI grain, it has a higher concentration in pyrrhotite. Octahedrally coordinated $Cu^{2+}$ has similar ionic radii to both $Fe^{2+}$ and $Ni^{2+}$ in the M1 and M2 sites in both minerals (percent differences of 6–7% in the M1 sites of both pyrrhotite and pentlandite and 4–10% in the M2 site of pentlandite). However, if tetrahedrally coordinated $Cu^{1+}$ is the most appropriate, it is essentially too large for the M1 sites in either pyrrhotite or pentlandite (percent differences of 14–26%) but similar to $Fe^{3+}$ for pyrrhotite (percent difference 8%) and $Ni^{2+}$ or $Fe^{2+}$ for the M2 site in pentlandite (percent differences of 9% and 5%, respectively). Given the similar ionic radii, it could be that the substitution of Cu into pyrrhotite depends on the amount of vacancies the pyrrhotite contains, and hence the iron deficiency of the pyrrhotite. Comparing the metal to sulfur+selenium ratios (calculated as Fe+Ni+Co+Cu+Ge+Zn / S+Se in atomic percent) of the pyrrhotite in the CM and CR PPI grains, the CM PPI pyrrhotite has slightly more Fe-deficiency (M/S+Se = 0.989) than the CR PPI pyrrhotite (M/S+Se = 0.986).

The higher Fe-deficiency is consistent with greater Cu contents in the CM PPI Po (870 ppm for the CM versus 198 ppm for the CR). Still, this only explains the relative differences in Cu abundance between the pyrrhotites, not the differences in the pyrrhotite and pentlandite partitioning behaviors. Investigating the role of electronegativity of the elements for Cu and Ni (1.90 and 1.91, respectively) versus Cu and Fe (1.90 and 1.83, respectively), we find that the percent difference is lower between Cu and Ni (0.52%) than between Cu and Fe (3.75%). In the case of Cu, however, the percent difference is even lower for Ni than was the case with Co. This ought to result in Cu preferentially substituting into pentlandite more than pyrrhotite, but that is not the case with the CM PPI grain. It could be that Cu substitution is dependent on both Fe-deficiency in pyrrhotites and electronegativity differences in pentlandite, with the relative effects determining which phase ends up incorporating more or less Cu. However, the solid-state unmixing behavior of Cu in pyrrhotite and pentlandites, as derived from first principles, does not appear to adequately explain the behavior observed from the physical samples. Instead, the cooling rates of the grains in question likely had the greatest effect on the Cu partitioning as discussed below.

Comparing CM and CR PPI partitioning behavior shows, for all elements except Ge, stronger partitioning into the CR PPI grain. In other words, quantitatively, the difference between $D_i$ and 1 is greatest or, qualitatively, the magnitude of the height of the bar in Figure 10 is largest for the CR PPI grain. Partitioning can occur under two scenarios: 1) equilibrium partitioning, where time is not important and phase separation is solely a function of temperature, and 2) kinetically-controlled partitioning, where time, and hence the amount of diffusion, controls the degree of phase separation and the resulting partition coefficient. In the former case, any differences in element distributions between PPI grains would be a function of the equilibration temperature, whereas in the latter case, differences would be a function of the cooling rate and, consequently, be dependent on time. Diffusion rates of metals in sulfides are rapid and so it is possible that for slower chondrule cooling rates, equilibrium partitioning may have been maintained for the higher temperature part of the cooling path. However, at lower temperatures, it seems most likely that partitioning would depart from equilibrium, and for fast cooling rates, this would be the norm. Therefore, kinetically-controlled partitioning is probably most likely. In this case, the extent of the partitioning is a function of the degree to which the pyrrhotite and pentlandite unmix from *mss,* which itself is dependent on the cooling rate. For grains which cooled slowly down to low temperatures, the partitioning between pyrrhotite and pentlandite is greater, because there was more time for phase separation and diffusion to take place. If this is the case, we would expect that for PPI grains, which experienced greater phase separation, there should be both larger $D_i$ and also smaller concentrations of Ni in the pyrrhotite. This is indeed what is observed in the CR PPI grain, which contains 7,740 ppm Ni in pyrrhotite (compared to 12,500 ppm Ni in the pyrrhotite of the CM PPI grain; Table 2). Relatedly, if Cu partitions into pentlandite, a grain with slower cooling rate would have a higher concentration of Cu in pentlandite (what we see in the CR PPI grain) and one with a faster cooling rate would have a higher concentration of Cu in the pyrrhotites (what we see in the CM PPI grain). The cooling rate is also a function of chondrule size, with larger chondrules having slower cooling rates. The CR PPI grain is present in a larger chondrule (610 × 810 µm) as compared to the CM PPI grain's chondrule (310 × 610 µm). This larger size and potentially slower cooling rate are consistent with the CR PPI grain having larger $D_i$, smaller concentrations of Ni in the pyrrhotite, and higher concentrations of Cu in the pentlandite,

However, when discussing element concentrations in pyrrhotite, we must also consider the presence of submicron pentlandite inclusions in the pyrrhotite as we observed in Figure 4. This implies that both a major, high temperature episode of phase separation occurred which produced the large pentlandite exsolution textures, and, subsequently, a lower temperature stage of exsolution occurred, which produced the submicron pentlandite. It is important to point out that the analytical volume of the SXRF analyses of the PPI grain pyrrhotite would have included the fine-scale pentlandite. This means our measured trace element compositions represent the high temperature episode of phase separation. With this in mind, the Ni contents of the pyrrhotite, and contents of all other elements that partition into pentlandite, are even lower than our SXRF analyses show; therefore, the $D_i$ values we calculate represent lower limits.

Lastly, a comparison of our trace elemental abundance data of the PPI grain sulfides with previous work could provide additional insights into the grains' histories, as well as information on how different analytical techniques compare in their ability to measure trace elements. Unfortunately, very few studies have looked at the trace elemental abundances of chondritic iron sulfides. The only studies which analyzed iron sulfides in CM or CR chondrites for the elements of interest are Dyl et al. (2014) and Jacquet et al. (2013). The former used quantitative SXRF mapping to characterize the distribution of Zn in and around iron sulfides in CM Murchison. They found Zn-rich (0.6–1.4 wt.%) rims around grains with textural features consistent with alteration, specifically altered PPI grains similar to those discussed in detail by Singerling and Brearley (2020). Interestingly, Dyl et al. (2014) observed the highest concentration of Zn in a rim around a pentlandite grain. This is consistent with our finding that Zn is preferentially incorporated into pentlandite. The rimming is likely the product of Zn mobilization during parent body aqueous alteration, given that Murchison is a moderately altered sample (2.5 on the scale developed by Rubin et al., 2007). The Jacquet et al. (2013) study included an LA-ICP-MS analysis of one sulfide grain rimming a metal. This grain is texturally more similar to an SRM grain and, as such, will be discussed in the following section.

Trace elemental abundances have been measured for iron sulfides in other chondrite groups, including using SXRF on CV3 chondrites (Brearley, 2007) and CI Orgueil (Greshake et al., 1998), as well as LA-ICP-MS on LL5 Chelyabinsk (Andronikov et al., 2015) and G chondrite Sierra Gorda 009 (Ivanova et al., 2020). However, these different meteorite groups represent samples from different parent bodies with distinct histories, and so a comparison between the trace elemental abundances of their iron sulfides to those in the current study is unlikely to provide meaningful information. Additionally, the small grain size of sulfides in carbonaceous chondrites has historically been an impediment to carrying out large numbers of trace elemental analyses by LA-ICP-MS, which explains the current lack of data. Although SIMS has been used to study chondritic sulfides, such works primarily focus on the isotopic compositions of the sulfides rather than the trace elemental abundances. The limited data currently available for elements such as Cu, Ge, Zn, and Se from iron sulfides and associated metal in CM and CR chondrites further highlights the significance of the current analyses. Future studies will hopefully fill in the gaps in the dataset.

### 4.2. SRM Grain Trace Element Patterns and Abundances

The SRM grains are postulated to have formed in the solar nebula as well (e.g., Schrader and Lauretta, 2010; Schrader et al., 2015; Singerling and Brearley, 2018). However, since the SRM grains in CR chondrites only occur in chondrules, dominantly type IA, the formation of the precursor metal must involve a formational mechanism other than simple nebular condensation.

The following is a brief summary of the processes involved in the formation of these sulfide-metal assemblages, what we refer to as the sulfidization model:

1) Aggregation of dust consisting of silicate, metal, and (perhaps) sulfide precursors surrounded by gases in the solar nebula.
2) Formation of chondrules by flash heating, with loss of volatile elements.
3) Formation of immiscible silicate and metal±sulfide melts within the type IA chondrules.
4) Migration of immiscible metal±sulfide melts from the chondrule interiors to the rims from spinning and acceleration of the chondrule (Grossman and Wasson, 1985; Connolly et al., 1994; Tsuchiyama et al., 2000); the high volatility of sulfur could have caused it to escape from the chondrules, if it was initially present, into the surrounding gas and form $H_2S$.
5) Crystallization of Fe,Ni metal on chondrule rims, as well as crystallization of silicate phases in the chondrule interiors.
6) Sulfidization of Fe,Ni metal from reaction with surrounding $H_2S$ gas to form *mss*; depending on the temperature and duration of the reaction, elements, such as Ni, could diffuse back into the metal resulting in different element ratios between the sulfide and metal.
7) Depending on the cooling rate of the system, *mss* could subsequently exsolve pentlandite and transform into pyrrhotite.

This model is similar to that proposed by Zanda et al. (2002). As is apparent above, the SRM grains do not represent primary condensate compositions in the canonical sense (i.e., condensates with little subsequent processing); however, they do represent compositions of the solar nebula as modified by at least some of the aforementioned processes.

The overall patterns of the SRM pyrrhotite and metal are strikingly similar, with the actual abundances of individual elements, relative to CI chondrite, varying between the two phases (Fig. 7e). This implies a genetic link between the two, such as sulfidization. Both pyrrhotite and metal show enrichments in Co and Cu, consistent with the chalcophile-siderophile nature of these elements. Selenium, on the other hand, is enriched in the pyrrhotite and depleted in the metal. If the pyrrhotite formed from the metal via sulfidization, the Se must have been inherited from either the metal, which is depleted in Se, or from the gas.

Both Ge and Zn are depleted in the pyrrhotite and metal. The depletion of Zn in the SRM pyrrhotite is entirely consistent with the idea that metal is the precursor phase, because it has exactly the same Zn depletion as the pyrrhotite. The depletion of Zn in metal is to be expected, because Zn is chalcophile in nature and type I chondrules in CR chondrites are even depleted in elements that condense at higher temperatures (e.g. Au, Na, Rb; Connolly et al., 2001). Germanium displays complex behavior; it is more depleted in SRM pyrrhotite than is observed in PPI pyrrhotite and is depleted in SRM metal even though it has a siderophile nature. We next consider if the proposed processes responsible for the formation of these assemblages can explain the observed complexities in the trace element concentrations.

Significant processes that likely contributed to the formation of the SRM grains include: volatilization during chondrule formation, formation of immiscible silicate and metal±sulfide melts (liquid-liquid partitioning), mobilization of the metal±sulfide melts to chondrule rims, crystallization of silicate solids and metal (liquid-solid partitioning), sulfidization of metal, and small amounts of unmixing into pyrrhotite and pentlandite (solid-solid partitioning). Chondrule formation could have potentially removed some proportion of the more volatile elements (Ge,

Zn, Se) from chondrules. As mentioned previously, the lack of sulfides, as well as lithophile volatile elements, in type I chondrules implies either that these chondrules did not retain their more volatile elements or never contained any to begin with.

Immiscible melt formation would have partitioned elements between silicate and metal±sulfide fractions in the type I chondrules. Nickel, Co, Cu, and Ge should partition into the metal melt, exactly as we observe in metal in the SRM grains. However, Ge is depleted suggesting that it has been lost from the chondrule. Germanium has a low condensation temperature (883 K; Lodders, 2003) and a relatively high diffusion coefficient (Righter et al., 2005). It could have been volatilized along with S during chondrule formation, which includes migration of metal melts to chondrule rims. However, unlike S, it did not recondense onto/into the metal or in the sulfide to a large extent and instead reacted with some other component in the nebular gas/dust. Alternatively, condensation temperatures of precursor materials could have played a role. While Ge is cosmochemically siderophile, it is depleted in the SRM metal. This could be due to the low condensation temperature of Ge (50% $T_C$ of 883 K) as compared to the other siderophile elements listed, as well as Fe alloy (50% $T_C$ of 1353, 1352, and 1037 K for Ni, Co, and Cu, respectively and $T_C$ of 1357 K for Fe alloy) (Lodders et al., 2003). Therefore, although Ge may be siderophile, its low condensation temperature precluded metal condensates having large Ge concentrations. That is, condensate Fe,Ni alloys would have largely already formed before Ge was able to condense into them. Any subsequent processing of metal condensates, such as melting during chondrule formation, would be similarly depleted in Ge.

Alternatively, Ge could have escaped volatilization and remained in the metal, given its siderophile nature. Its high diffusivity may have caused it to migrate to the metal-sulfide boundary as sulfidization proceeded, where it accumulated and formed small (nm-scale) metallic residues enriched in Ge. In fact, Singerling and Brearley (2018) observed the presence of Ni(P)-rich metal, phosphate, chromite, and pentlandite along the metal-sulfide interfaces of three SRM grains (one from CM chondrite and two from CR chondrites). The geometry of our two FIB sections from the SRM grain in the current study do not include the metal-sulfide interface (Fig. 2f) but, rather, are oriented parallel to it, so it is possible that Ge is present but was simply not sampled in our analyses. Future TEM studies of metal-sulfide boundaries in SRM grains would help resolve this question.

The key difference between type I and II chondrules is the absence or presence of sulfur which, in turn, results in different immiscible melts (i.e., silicate and metal for type I chondrules versus silicate and *mss* for type II chondrules). Zinc partitions both into the silicate melt and into the sulfide melt (MacLean and Shimazaki, 1976; Shimazaki and MacLean, 1976); therefore, in the type II chondrules it likely partitioned into both melts (silicate and *mss*), whereas in the type I chondrules it likely partitioned only into the silicate melt since it has no siderophile affinities. This implies that while Zn would be detectable in the products of the *mss* melt (i.e., the PPI grains), it would not be present in large concentrations in the products of the metal melt (i.e., the SRM grains), having gone into the silicate portions of the type I chondrules instead. As a consequence, low concentrations of Zn are expected, and observed in both metal and sulfide in the SRM grains.

Sulfidization of the Fe,Ni metal would have partitioned elements with chalcophile affinities from the metal into the *mss* that formed. This would result in a lower concentration of those elements in the Fe,Ni metal compared to the precursor metal. Additionally, the *mss* that formed from the sulfidization could have incorporated other volatile elements similar to S, such as Se, into its structure. This is consistent with Lodder's (2003) prediction that Se would condense into troilite. We observe enrichments of Se in the pyrrhotite of the SRM grains consistent with this mechanism. As sulfidization proceeds, the concentration of the precursor metal grain will be modified as elements repartition between the remaining metal and the newly-formed sulfide. The trace element composition of the metal will therefore evolve as sulfidization proceeds, provided temperatures are high enough to allow diffusion to occur, otherwise diffusion profiles will develop within the metal. This change in composition of the metal is a distinct process from the PPI grains, which started off with a fixed bulk composition that then experienced phase separation resulting in partitioning between the phases.

The element distribution (${}^{\alpha/\beta}D_i$) data for the SRM grain, summarized in Table 5 and Figure 11, quantifies the partitioning behavior between the metal and the sulfide. Formation under equilibrium conditions for the SRM grain is less likely than the case for the PPI grains, given the mechanism proposed by Zanda et al. (2002) and shown in Figure 8b. However, these element distribution data are not likely to be true partition coefficients but instead are presented for illustrative purposes to enable comparison with equilibrium partition coefficients. As there is no convention for calculation of ${}^{\alpha/\beta}D_i$ in solid phase-solid phase systems for sulfide-metal assemblages, we have calculated the values as metal/pyrrhotite (α/β) for the SRM grains. For $D_i$ >1, the element preferentially partitions into metal, and for $D_i$ <1, the element preferentially partitions into pyrrhotite. The key observation is that there is a linear correlation between element volatility and partitioning. As element volatility increases (moving to the right of Fig. 11), the $D_i$ decreases, illustrating increasing likelihood for incorporation of the element into pyrrhotite. This relationship between volatility and concentration is expected in sulfidization models; as metal grains cooled, they would increasingly cease to equilibrate with the nebular gas (Grossman and Wasson, 1985). The high temperatures present during chondrule formation promoted volatilization, and in the case of the type I chondrules, these volatile elements were not retained. Only at lower temperatures where sulfidization is predicted to have occurred would these elements be able to be incorporated into chondrules. Furthermore, the fact that Zn and Se are partially (Zn) to highly (Se) chalcophile obviously favors incorporation of both elements into pyrrhotite.

Lastly, as with the PPI grains, a comparison of our trace elemental abundance data for the SRM grain with previous works could provide useful information. Unlike the case with the iron sulfides, there have been several detailed studies of trace elements in CR chondrite metals (e.g., Connolly et al., 2001; Campbell et al., 2002; Humayun et al., 2002; Jacquet et al., 2013; Weyrauch et al., 2021). The SRM metal grain from this study (C3S3) has a Ni/Co ratio of 20.7, which is consistent with the solar ratio (21.3, Palme et al., 2014) and previous works—21.6 from the metals in CR EET 92042 (the same meteorite SRM grain C3S3 was found in) by Connolly et al. (2001); 22.0 from the metals in multiple CR chondrites by Jacquet et al. (2013); and 21.7 from the metals in CR NWA 852 by Weyrauch et al. (2021). A solar Ni/Co ratio for metals has been used to argue for formation via condensation (Weisberg et al., 1993); however, SRM grain C3S3 is clearly not a product of direct nebular condensation from a gas given its intimate association with a chondrule. Solar Ni/Co ratios of CR metals have also been explained by more extensive melting of chondrules (Zanda et al., 2002). In fact, the chondrule SRM grain C3S3 is located in is consistent with the model proposed by Zanda et al. (2002); the chondrule is nearly spherical in shape with few interior metal grains, a well-defined metal rim, and coarse-grained silicates.

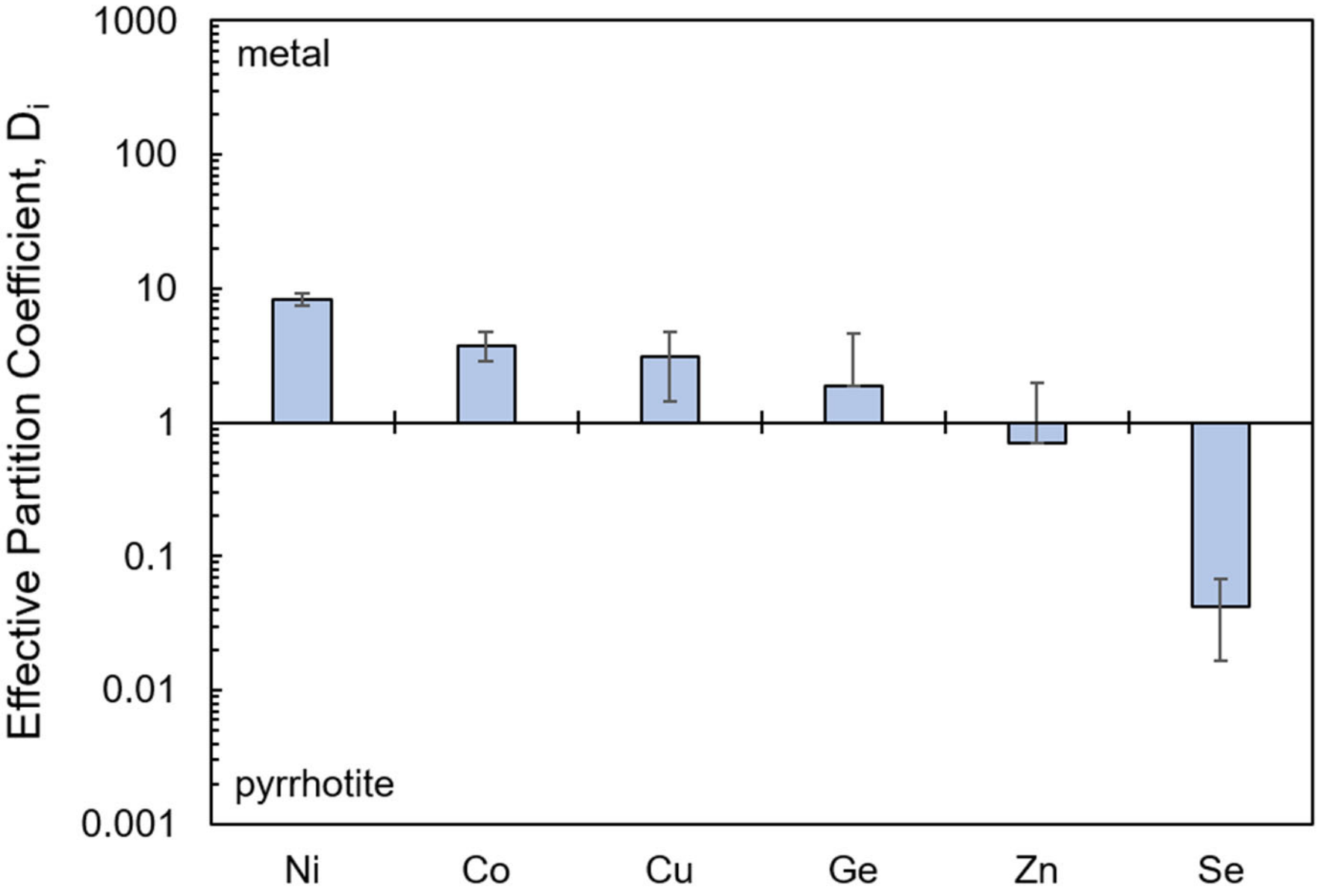

**Figure 11.** Bar graph illustrating the effective partition coefficients for metal v. pyrrhotite in the SRM grain. All elements except for Zn and Se are preferentially incorporated into metal rather than pyrrhotite. There is a correlation between element volatility and partitioning; moving towards the right (i.e., higher volatility), the elements are preferentially incorporated less into the metal and more into the pyrrhotite.

The Ge and Se abundances of SRM grain C3S3 (2.8 ppm Ge and 9.8 ppm Se) are also in agreement with previous works; the Ge in CR metals ranges from 0.6–5 ppm with an average value of 1.7 ppm (Weyrauch et al., 2021), and the Se in CR metals ranges from 2.9–37.0 ppm with an average value of 14 ppm (Jacquet et al., 2013). As mentioned previously, Jacquet et al. (2013) identified a metal grain rimmed by sulfide, essentially an SRM grain, and obtained trace element data for both the metal and sulfide portions of the assemblage, which they labeled g23. They found that the g23 metal contained 37 ppm Se, compared to the 9.8 ppm in SRM grain C3S3, and the g23 sulfide contained 86 ppm Se, compared to the 230 ppm Se in SRM grain C3S3. Although SRM grain C3S3 pyrrhotite has a higher abundance of Se compared to the g23 sulfide, the overall elemental distribution showing a preference for Se in the sulfide over metal portions of SRM grains is true for both SRM grain C3S3 and g23.

Unlike the aforementioned elements, the abundance of Cu in SRM grain C3S3 (870 ppm Cu) differs from previous studies. The average Cu abundance for CR metals is 55 and 45 ppm from Jacquet et al. (2013) and Weyrauch et al. (2021), respectively. Still, the Cu contents vary widely in CR metals, reaching as high as 465 ppm in SRM g23 metal (Jacquet et al., 2013). Additionally, the location of the metals within the meteorites appears to affect the Cu content, with isolated metals in the matrix and metals in the interior of chondrules having lower average Cu contents (36 and 43 ppm Cu, respectively) as compared to metals located on the rims of chondrules (average of 70 ppm Cu) (Jacquet et al., 2013). For SXRF analyses, the Ni Kβ peak can overlap with Cu Kα; therefore, it is sometimes preferable to use the Cu Kβ peak for fitting SXRF spectra to obtain Cu abundances. For pentlandite, we opted to use the Cu Kβ peak for peak fitting purposes; however, in the spectra from our low Ni phases (i.e., pyrrhotite and metal), the Cu Kα is well resolved from the Ni Kβ peak and so was used. Given the well-resolved peaks and our ability to obtain good peak fits, we argue that our data for Cu is reflective of the actual samples and not an analytical problem with SXRF analyses. Instead, we argue that the difference between the Cu abundances for the CR metal analyses between this study and those involving LA-ICP-MS is likely a product of the different spatial resolutions of the techniques used. Our SXRF analyses had a spot size of 2 µm, whereas those from Jacquet et al. (2013) were 51–77 µm and those from Weyrauch et al. (2021) were 30 µm. If there are very small-scale localized variations in the distribution of Cu, these would tend to be averaged out for the LA-ICP-MS analyses, resulting in a lower Cu content. Additional SXRF analyses of Cu in CR metal could help determine if the high Cu contents are a result of the higher resolution technique or else a feature of the specific SRM grain we analyzed.

### 4.3. Comparing PPI and SRM Trace Element Patterns and Abundances

Despite being from different meteorite groups and having formed by different mechanisms, the SRM and PPI grains show remarkably similar characteristics in terms of the shapes of their volatility-related abundance patterns and the relative abundances of the elements. The similarities between the PPI bulk compositions (*mss*), the SRM pyrrhotite, and the SRM metal (Figure 12) are likely due to similar processes involved in their formation (e.g., volatilization; *l-l*, *l-s*, and *s-s* partitioning). By comparing these patterns, we can establish if the observed patterns are consistent with our proposed formation mechanisms (i.e., crystallization model for the PPI grains and sulfidization model for the SRM grains). In short, the SRM pyrrhotite has smaller amounts of Ni, Co, and Cu than the PPI *mss*, because these elements partitioned more strongly into the metal and are largely retained in the metal even after sulfidization, whereas in the PPI grains, the elements partitioned into the *mss*. The SRM pyrrhotite has smaller amounts of Ge and Zn compared to PPI pyrrhotite owing to the different processes and conditions experienced by the type I and IIA chondrules—silicate-metal vs silicate-*mss* partitioning, retention of volatile elements, cosmochemical behavior of elements, and condensation temperatures of elements in phases of interest (metal, sulfide, silicate).

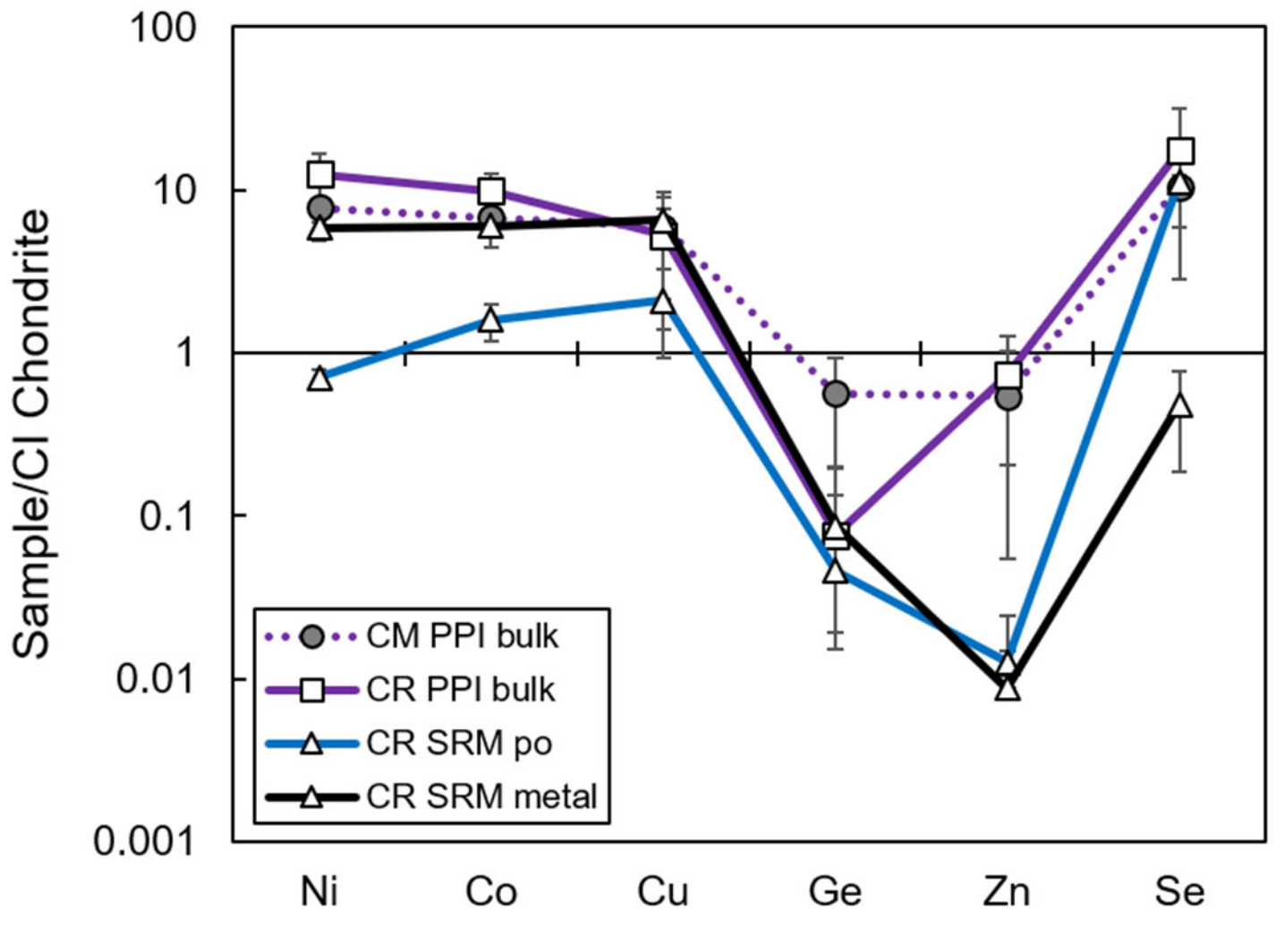

**Figure 12.** CI-normalized abundance plot, as described in Figure 6, comparing SXRF data for CR SRM metal and pyrrhotite to bulk compositions of the CM and CR PPI grains. Note the similarities between the SRM metal and pyrrhotite with the PPI bulk. Bulk refers to the bulk composition of the PPI, which we assume is equivalent to the composition of the *mss*. The line form corresponds to the meteorite group (CM = dashed, CR = solid), the symbol to the textural group (CM PPI = gray circle, CR PPI = white square, and SRM = white triangle), and the line color to the phase (Fe,Ni metal = black, pyrrhotite = blue, bulk = purple).

## 5. CONCLUSIONS

This work, the first of its kind to look at trace elements and microstructures in individual chondritic sulfide grains, involved SXRF microprobe analyses on FIB-prepared sections of pyrrhotite, pentlandite, and Fe,Ni metal from representative occurrences of primary sulfide and metal/sulfide grains in CM and CR carbonaceous chondrites. We detected Fe, Ni, Co, Cu, Ge, Zn, and Se. Several types of grains that we analyzed have remarkably similar trace element abundance patterns including: 1) CM and CR PPI grain pyrrhotite, 2) CM and CR PPI grain pentlandite, 3) CM and CR PPI bulk, and 4) SRM grain pyrrhotite and metal. The similarity between the CM and CR sulfides is further evidence for similar formation mechanisms and conditions. The similarity between the SRM pyrrhotite and metal supports a genetic relationship between the two phases that would expected as a result of sulfidization of metal.

The PPI grain bulk composition has enrichments in Ni, Co, Cu, and Se and depletions in Ge and Zn relative to CI chondrite values, a likely result of liquid-liquid partitioning between silicate and *mss* melts during chondrule formation. The PPI grain pentlandite has low concentrations of Ge and high concentrations of Zn, whereas the PPI grain pyrrhotite shows the opposite behavior. These are a likely result of solid-state unmixing between pyrrhotite and pentlandite from *mss*. The SRM metal has enrichments of Ni, Co, and Cu and depletions of Ge, Zn, and Se. The SRM pyrrhotite has enrichments of Co, Cu, and Se and depletions of Ni, Ge, and Zn. These are likely a result of the loss of volatile elements during chondrule formation, the cosmochemical behavior of the elements, and the condensation temperatures into the phases of interest. The $D_i$ values of the SRM grain show a linear correlation with element volatility for the metal-pyrrhotite system, which is the result of formation by sulfidization.

### Acknowledgements

Lindsay Keller, Rhian Jones, Elena Dobrica, Jon Lewis, and Mike Spilde provided much appreciated advice and recommendations for this work. We thank Associate Editor Pierre Beck and reviewers Dennis Harries, Herbert Palme, and an anonymous reviewer for excellent feedback and suggestions for improving the manuscript. We acknowledge the Meteorite Working Group for providing the samples used in this study. U.S. Antarctic meteorite samples are recovered by the Antarctic Search for Meteorites (ANSMET) program, which has been funded by NSF and NASA, and characterized and curated by the Department of Mineral Sciences, the Smithsonian Institution and Astromaterials Curation Office at NASA Johnson Space Center. Electron microscopy and FIB sample preparation were carried out in the Electron Microbeam Analysis Facility in the Department of Earth and Planetary Sciences, University of New Mexico, a facility supported by the State of New Mexico, NSF, and NASA. This research used resources of the Advanced Photon Source, a U.S. Department of Energy (DOE) Office of Science User Facility operated for the DOE Office of Science by Argonne National Laboratory under Contract No. DE-AC02-06CH11357. We acknowledge the support of GeoSoilEnviroCARS (The University of Chicago, Sector 13), at the Advanced Photon Source. GeoSoilEnviroCARS is supported by the National Science Foundation – Earth Sciences (EAR-1634415). This work was supported by NASA Cosmochemistry grants NNX11AK51G and NNX15AD28G to A.J. Brearley (PI).

## References

Andronikov A. V., Andronikova I. E., and Hill D. H. (2015) Impact history of the Chelyabinsk meteorite: Electron microprobe and LA-ICP-MS study of sulfides and metals. *Planet. Space Sci.* **118**, 54–78.

Barnes S.-J., Cox R. A., and Zientek M. L. (2006) Platinum-group element, gold, silver and base metal distribution in compositionally zoned sulfide droplets from the Medvezky Creek Mine, Noril'sk, Russia. *Contrib. Min. Pet.* **152**, 187–200.

Berlin J. (2009) Mineralogy and bulk chemistry of chondrules and matrix in petrologic type 3 chondrites: Implications for early solar system processes. Ph. D. thesis, Univ. of New Mexico.

Bernstein L. R. (1985) Germanium geochemistry and mineralogy. *Geochim. Cosmochim. Acta* **49**, 2409–2422.

Brearley A. J. (2007) Distribution of trace elements in sulfide and metal in reduced and oxidized CV3 carbonaceous chondrites determined by EPMA and SXRF. *Meteor. Planet. Sci.* **70**. #5239(abstr.).

Campbell A. J., Humayun M., and Zanda B. (2002) Partial condensation of volatile elements in Renazzo chondrules. *Geochim. Cosmochim. Acta* **66**(15A), A117.

Clementi E., Raimondi D. L., and Reinhardt W. P. (1963) Atomic Screening Constants from SCF Functions. *J. Chem. Phys.* **38**, 2686–2689.

Connolly H. C., Jr., Hewins R. H., Ash R. D., Zanda B., Lofgren G. E., and Bourot-Denise J. (1994) Carbon and the formation of reduced chondrules. *Nature* **371**, 136–139.

Connolly H. C., Jr., Huss G. R., and Wasserburg G. J. (2001) On the formation of Fe-Ni metal in Renazzo-like carbonaceous chondrites. *Geochim. Cosmochim. Acta* **65**, 4567–4588.

Cook N. J., Ciobanu C. L., Pring A., Skinner W., Shimizu M., Danyushevsky L., Saini-Eidukat B., and Melcher F. (2009) Trace and minor elements in sphalerite: A LA-ICPMS study. *Geochim. Cosmochim. Acta* **73**, 4761–4791.

Cook N. J., Etschmann B., Ciobanu C. L., Geraki K., Howard D. L., Williams T., Rae N., Pring A., Chen G., Johannessen B., and Brugger J. (2015) Distribution and substitution mechanism of Ge in a Ge-(Fe)-Bearing sphalerite. *Minerals* **5**, 117–132.

Criss J. W., Birks L. S., and Gilfrich J. V. (1978). Versatile X-ray analysis program combining fundamental parameters and empirical corrections. *Anal. Chem.* **50**, 33–37.

Dare S. A. S., Barnes S.-J., and Beaudoin G. (2012) Variation in trace element content of magnetite crystallized from a fractionating sulfide liquid, Sudbury, Canada: Implications for provenance discrimination. *Geochim. Cosmochim. Acta* **88**, 27–50.

Donovan J. J., Snyder D. A., and Rivers M. L. (1993) An improved interference correction for trace element analysis. *Microbeam Anal.* **2**, 23–28.

Dyl K. A., Bland P. A., and Cleverley J. S. (2014) Unraveling the record of nebular and asteroidal processes in the Murchison (CM2) meteorite: Pentlandite formation and its association with zinc mobility. *Lunar Planet. Sci. IXV.* Lunar Planet. Inst., Houston. #2311(abstr.).

Fleet M. E., Liu M., and Crocket J. H. (1999) Partitioning of trace amounts of highly siderophile elements in the Fe-Ni-S system and their fractionation in nature. *Geochim. Cosmochim. Acta* **63**, 2611–2622.

Flynn G. J. and Sutton S. R. (1992) Trace elements in chondritic stratospheric particles: Zinc depletion as a possible indicator of atmospheric entry heating. *Proc. Lunar Planet. Sci.* **22**, 171–184.

Flynn G. J., Sutton S. R., and Lanzirotti A. (2000) A comparison of the selenium contents of sulfides from interplanetary dust particles and meteorites. *Meteor. Planet. Sci.* **63**. #5265(abstr.).

Gaetani G. A. and Grove T. L. (1997) Partitioning of moderately siderophile elements among olivine, silicate melt, and sulfide melt: Constraints on core formation in the Earth and Mars. *Geochim. Cosmochim. Acta* **61**, 1829–1846.

Greshake A., Klock W., Arndt P., Maetz M., Flynn G. J., Bajt S., and Bischoff A. (1998) Heating experiments simulating atmospheric entry heating of micrometeorites: Clues to their parent body sources. *Meteor. Planet. Sci.* **33**, 267–290.

Grossman J. N. and Wasson J. T. (1985) The origin and history of the metal and sulfide components of chondrules. *Geochim. Cosmochim. Acta* **49**, 925–939.

Harries D. and Langenhorst F. (2013) The nanoscale mineralogy of Fe,Ni sulfides in pristine and metamorphosed CM and CM/CI-like chondrites: Tapping a petrogenetic record. *Meteor. Planet. Sci.* **48**, 879–903.

Helmy H. M., Ballhaus C., Wohlgemuth-Ueberwasser C., Fonseca R. O. C., and Laurenz V. (2010) Partitioning of Se, As, Sb, Te and Bi between monosulfide solid solution and sulfide melt – Application to magmatic sulfide deposits. *Geochim. Cosmochim. Acta* **74**, 6174–6179.

Humayun M., Campbell A. J., Zanda B., and Bourot-Denise M. (2002) Formation of Renazzo chondrule metal inferred from siderophile elements. *Lunar Planet. Sci. XXXIII.* Lunar Planet. Inst., Houston. #1965(abstr.).

Ivanova M. A., Lorenz C. A., Humayun M., Corrigan C. M., Ludwig T., Trieloff M., Righter K., Franchi I. A., Verchosky A. B., Korochantseva E. V., Kozlov V. V., Teplyakova S. N., Korochantsev A. V., and Grokhovsky V. I. (2020) Sierra Gorda 009: A new member of the metal-rich G chondrites grouplet. *Meteor. Planet. Sci.* **8**, 1764–1792.

Jacquet E., Paulhiac-Pison M., Alard O., Kearsley A. T., and Gounelle M. (2013) Trace element geochemistry of CR chondrite metal. *Meteor. Planet. Sci.* **48**, 1981–1999.

Kallemeyn G. W., Rubin A. E., Wang D., and Wasson J. T. (1989) Ordinary chondrites: Bulk compositions, classification, lithophile-element fractionations, and composition-petrographic type relationships. *Geochim. Cosmochim. Acta* **53**, 2727–2767.

Kimura M., Grossman J. N., and Weisberg M. K. (2011) Fe-Ni metal and sulfide minerals in CM chondrites: An indicator for thermal history. *Meteor. Planet. Sci.* **46**, 431–442.

Kitakaze A., Sugaki A., Itoh H., and Komatsu R. (2011) A revision of phase relations in the system Fe-Ni-S from 650°C to 450°C. *Can. Min.* **49**, 1687–1710.

Kullerud G. (1963) The Fe-Ni-S system. *Carn. Instit. Wash. Year Book* **62**, 175–189.

Large R. R., Danyushevsky L., Hollit C., Maslennikov V., Meffre S., Gilbert S., Bull S., Scott R., Emsbo P., Thomas H., Singh B., and Foster J. (2009) Gold and trace element zonation in pyrite using a laser imaging technique: Implications for the timing of gold in orogenic and Carlin-style sediment-hosted deposits. *Econ. Geol.* **104**, 635–668.

Li C., Barnes S.-J., Makovicky E., Rose-Hansen J., and Makovicky M. (1996) Partitioning of nickel, copper, iridium, rhenium, platinum, and palladium between monosulfide solid solution and sulfide liquid: Effects of composition and temperature. *Geochim. Cosmochim. Acta* **60**, 1231–1238.

Liu M. and Fleet M. E. (2001) Partitioning of siderophile elements (W, Mo, As, Ag, Ge, Ga, and Sn) and Si in the Fe-S system and their fractionation in iron meteorites. *Geochim. Cosmochim. Acta* **65**, 671–682.

Lodders K. (2003) Solar system abundances and condensation temperatures of the elements. *Astrophys. J* **591**, 1220–1247.
MacLean W. H. and Shimazaki H. (1976) The partition of Co, Ni, Cu, and Zn between sulfide and silicate liquids. *Econ. Geol.* **71**, 1049–1057.
Makovicky E. (2006) Crystal structures of sulfides and other chalcogenides. *Rev. Min. Geochem.* **61**, 7–125.
McSween H. Y., Jr. and Huss G. R. (2010) *Cosmochemistry*. Cambridge Univ. Press, Cambridge.
Paktunc A. D., Hulbert L. J., and Harris D. C. (1990) Partitioning of the platinum-group and other trace elements in sulfides from the Bushveld Complex and Canadian occurrences f nickel-copper sulfides. *Can. Min.* **28**, 475–488.
Palme H., Lodders K., and Jones A. (2014) Solar system abundances of the elements. *Planet. Aster. Com. Sol. Sys.* **1**, 15–36.
Papike J. J., Burger P. V., Shearer C. K., Sutton S. R., Newville M., Choi Y., and Lanzirotti A. (2011) Sulfides from martian and lunar basalts: Comparative chemistry for Ni, Co, Cu, and Se. *Am. Min.* **96**, 932–935.
Righter K., Campbell A. J., and Humayun M. (2005) Diffusion of trace elements in FeNi metal: Application to zoned metal grains in chondrites. *Geochim. Cosmochim. Acta* **69**, 3145–3158.
Rubin A. E., Trigo-Rodriguez J. M., Huber H., and Wasson J. T. (2007) Progressive aqueous alteration of CM carbonaceous chondrites. *Geochim. Cosmochim. Acta* **71**, 2361–2382.
Sahlström F., Arribas A., Dirks P., Corral I., and Chang Z. (2017) Mineralogical distribution of germanium, gallium, and indium at the Mt Carlton high-sulfidation epithermal deposit, NE Australia, and comparison with similar deposits worldwide. *Minerals* **7**, 213.
Schmitt W., Palme H., and Wanke H. (1989) Experimental determination of metal/silicate partition coefficients for P, Co, Ni, Cu, Ga, Ge, Mo, and W and some implications for the early evolution of the Earth. *Geochim. Cosmochim. Acta* **53**, 173–185.
Schrader D. L. and Lauretta D. S. (2010) High-temperature experimental analogs of primitive meteoritic metal-sulfide oxide assemblages. *Geochim. Cosmochim. Acta* **74**,1719–1733.
Schrader D. L., Connolly H. C., Jr., Lauretta D. S., Zega T. J., Davidson J., and Domanik K. J. (2015) The formation and alteration of the Renazzo-like carbonaceous chondrites III: Toward understanding the genesis of ferromagnesian chondrules. *Meteor. Planet. Sci.* **50**, 15–50.
Schrader D. L., Davidson J., and McCoy T. J. (2016) Widespread evidence for high-temperature formation of pentlandite in chondrites. *Geochim. Cosmochim. Acta* **189**, 359–376.
Shannon R. D. (1976) Revised effective ionic radii and systematic studies of interatomic distances in halides and chalcogenides. *Acta Crystallographica* **A32**, 751–767.
Shimazaki H. and MacLean W. H. (1976) An experimental study on the partition of zinc and lead between the silicate and sulfide liquids. *Min. Dep.* **11**, 125–132.
Singerling S. A. (2018) Primary and altered iron sulfides in CM and CR carbonaceous chondrites: Insights into nebular and parent body processes. Ph. D. thesis, Univ. of New Mexico.
Singerling S. A. and Brearley A. J. (2018) Primary iron sulfides in CM and CR carbonaceous chondrites: Insights into nebular processes. *Meteor. Planet. Sci.* **53**, 2078–2106.

Singerling S. A. and Brearley A. J. (2020) Altered primary iron sulfides in CM2 and CR2 Carbonaceous chondrites: Insights into parent body processes. *Meteor. Planet. Sci.* **55**, 496–523.
Tsuchiyama A., Kawabata T., Kondo M., Uesugi K., Nakano T., Suzuki Y., Yagi M., Umetani K., and Shirono S. (2000) Spinning chondrules deduced from their three-dimensional structures by X-ray CT method. *Lunar Planet. Sci. XXXI.* Lunar Planet. Inst., Houston. #1566(abstr.).
van Acken D., Humayun M., Brandon A. D., and Peslier A. H. (2012) Siderophile trace elements in metals and sulfides in enstatite achondrites record planetary differentiation in an enstatite chondritic parent body. *Geochim. Cosmochim. Acta* **83**, 272–291.
Wasson J. T. and Kallemeyn G. W. (1988) Compositions of chondrites. *Phil. Tran. Roy. Soc. A: MPES* **325**, 535–544.
Weisberg M. K., Prinz M., Clayton R. N., and Mayeda T. K. (1993) The CR (Renazzo-type) carbonaceous chondrite group and its implications. *Geochim. Cosmochim. Acta* **57**, 1567–1586.
Weyrauch M., Zipfel J., and Weyer S. (2021) The relationship of CH, CB, and CR chondrites: Constraints from trace elements and Fe-Ni isotope systematics in metals. *Geochim. Cosmochim. Acta* **308**, 291–309.
Yu Y., Hewins R. H., and Zanda B. (1996) Sodium and sulfur in chondrules: Heating time and cooling curves. In *Chondrules and the Protoplanetary Disk* (eds. R. H. Hewins, R. H. Jones, and E. R. D. Scott). Cambridge Univ. Press, Cambridge. pp. 213–219.
Zanda B., Bourot-Denise M., Hewins R. H., Cohen B. A., Delaney J. S., Humayun M., and Campbell A. J. (2002) Accretion textures, iron evaporation and re-condensation in Renazzo chondrules. *Lunar Planet. Sci. XXXIII.* Lunar Planet. Inst., Houston. #1852(abstr.).

**Table A1.** EPMA WDS and TEM EDS analyses

| Sample | Phase | Method | Point | Fe | S | Ni | Co | Cr | Total | Fe | S | Ni | Co | Cr | Total |
|---|---|---|---|---|---|---|---|---|---|---|---|---|---|---|---|
| | | | | (wt%) | | | | | | (at%) | | | | | |
| QUE 97990 (CM2, PPI, C12S2) | Po | EPMA WDS | 1 | 62.01 | 36.63 | 1.25 | 0.19 | 0.02 | 100.11 | 48.74 | 50.16 | 0.94 | 0.14 | 0.02 | 100.00 |
| | | | 2 | 61.76 | 36.95 | 1.37 | 0.19 | 0.03 | 100.29 | 48.39 | 50.43 | 1.02 | 0.14 | 0.02 | 100.00 |
| | | | 3 | 61.57 | 36.78 | 1.42 | 0.21 | 0.02 | 100.00 | 48.40 | 50.37 | 1.06 | 0.16 | 0.02 | 100.00 |
| | | | 4 | 62.00 | 36.77 | 1.10 | 0.17 | 0.03 | 100.07 | 48.71 | 50.31 | 0.82 | 0.12 | 0.03 | 100.00 |
| | | | 5 | 61.59 | 36.87 | 1.46 | 0.20 | 0.04 | 100.16 | 48.33 | 50.40 | 1.09 | 0.15 | 0.04 | 100.00 |
| | | | 6 | 61.82 | 36.53 | 1.01 | 0.19 | 0.05 | 99.60 | 48.82 | 50.24 | 0.76 | 0.14 | 0.04 | 100.00 |
| | | | 7 | 61.87 | 36.83 | 1.28 | 0.17 | 0.05 | 100.21 | 48.54 | 50.34 | 0.95 | 0.13 | 0.04 | 100.00 |
| | | | Average | 61.80 | 36.77 | 1.27 | 0.19 | 0.04 | 100.06 | 48.56 | 50.32 | 0.95 | 0.14 | 0.03 | 100.00 |
| | | | Std Dev | 0.18 | 0.14 | 0.18 | 0.02 | 0.01 | 0.25 | 0.19 | 0.07 | 0.13 | 0.01 | 0.01 | 0.00 |
| | Pn | EPMA WDS | 1 | 40.47 | 33.20 | 24.77 | 0.68 | 0.02 | 99.14 | 33.03 | 47.20 | 19.23 | 0.53 | 0.02 | 100.00 |
| | | | 2 | 40.38 | 33.56 | 24.64 | 0.72 | 0.02 | 99.31 | 32.83 | 47.53 | 19.06 | 0.56 | 0.01 | 100.00 |
| | | | Average | 40.42 | 33.38 | 24.70 | 0.70 | 0.02 | 99.23 | 32.93 | 47.37 | 19.15 | 0.54 | 0.02 | 100.00 |
| | | | Std Dev | 0.07 | 0.25 | 0.09 | 0.03 | 0.00 | 0.12 | 0.14 | 0.24 | 0.12 | 0.02 | 0.00 | 0.00 |
| QUE 99177 (CR2, PPI, C1S2) | Po | EPMA WDS | 1 | 62.49 | 36.91 | 0.84 | 0.14 | 0.03 | 100.45 | 48.92 | 50.33 | 0.63 | 0.10 | 0.02 | 100.00 |
| | | TEM EDS[1] | 1 | 58.97 | 40.96 | 0.08 | 0.00 | 0.00 | 100.00 | 45.23 | 54.72 | 0.06 | 0.00 | 0.00 | 100.00 |
| | | | 2 | 58.63 | 41.28 | 0.09 | 0.00 | 0.00 | 100.00 | 44.89 | 55.05 | 0.07 | 0.00 | 0.00 | 100.00 |
| | | | 3 | 54.81 | 41.05 | 0.07 | 0.00 | 0.00 | 100.00 | 45.12 | 54.81 | 0.07 | 0.00 | 0.00 | 100.00 |
| | | | Average | 57.47 | 41.10 | 0.08 | 0.00 | 0.00 | 100.00 | 45.08 | 54.86 | 0.07 | 0.00 | 0.00 | 100.00 |
| | | | Std Dev | 2.31 | 0.17 | 0.01 | 0.00 | 0.00 | 0.00 | 0.17 | 0.17 | 0.01 | 0.00 | 0.00 | 0.00 |
| | Pn | EPMA WDS | 1 | 39.64 | 33.76 | 24.21 | 0.80 | 0.02 | 98.70 | 32.42 | 48.10 | 18.84 | 0.62 | 0.02 | 100.00 |
| | | TEM EDS[1] | 1 | 37.98 | 38.37 | 23.65 | 0.00 | 0.00 | 100.00 | 29.83 | 52.50 | 17.67 | 0.00 | 0.00 | 100.00 |
| | | | 2 | 37.58 | 39.72 | 22.70 | 0.00 | 0.00 | 100.00 | 29.28 | 53.90 | 16.82 | 0.00 | 0.00 | 100.00 |
| | | | 3 | 36.13 | 39.90 | 23.97 | 0.00 | 0.00 | 100.00 | 28.13 | 54.11 | 17.76 | 0.00 | 0.00 | 100.00 |
| | | | Average | 37.23 | 39.33 | 23.44 | 0.00 | 0.00 | 100.00 | 29.08 | 53.50 | 17.42 | 0.00 | 0.00 | 100.00 |
| | | | Std Dev | 0.97 | 0.84 | 0.66 | 0.00 | 0.00 | 0.00 | 0.87 | 0.88 | 0.52 | 0.00 | 0.00 | 0.00 |
| EET 92042 (CR2, SRM, C3S3) | Po | EPMA WDS | 1 | 62.60 | 36.07 | 0.80 | 0.10 | 0.06 | 99.64 | 49.54 | 49.73 | 0.60 | 0.07 | 0.05 | 100.00 |
| | | | 2 | 62.82 | 35.89 | 0.73 | 0.08 | 0.05 | 99.56 | 49.80 | 49.56 | 0.55 | 0.06 | 0.04 | 100.00 |
| | | | 3 | 62.62 | 35.81 | 0.85 | 0.09 | 0.07 | 99.45 | 49.71 | 49.52 | 0.64 | 0.07 | 0.06 | 100.00 |
| | | | 4 | 62.91 | 36.11 | 0.67 | 0.06 | 0.06 | 99.82 | 49.70 | 49.69 | 0.51 | 0.04 | 0.05 | 100.00 |
| | | | Average | 62.74 | 35.97 | 0.76 | 0.08 | 0.06 | 99.62 | 49.69 | 49.62 | 0.57 | 0.06 | 0.05 | 100.00 |
| | | | Std Dev | 0.15 | 0.14 | 0.08 | 0.02 | 0.01 | 0.16 | 0.11 | 0.10 | 0.06 | 0.01 | 0.01 | 0.00 |
| | M | EPMA WDS | 1 | 92.92 | 0.01 | 6.31 | 0.30 | 0.07 | 99.74 | 93.57 | 0.02 | 6.04 | 0.29 | 0.08 | 100.00 |
| | | | 2 | 92.53 | 0.02 | 6.33 | 0.32 | 0.06 | 99.38 | 93.50 | 0.04 | 6.08 | 0.31 | 0.07 | 100.00 |
| | | | 3 | 92.42 | 0.00 | 6.50 | 0.31 | 0.09 | 99.43 | 93.35 | 0.00 | 6.25 | 0.30 | 0.10 | 100.00 |
| | | | 4 | 92.67 | 0.01 | 6.33 | 0.30 | 0.07 | 99.47 | 93.53 | 0.03 | 6.08 | 0.28 | 0.08 | 100.00 |
| | | | Average | 92.63 | 0.01 | 6.37 | 0.31 | 0.08 | 99.51 | 93.49 | 0.02 | 6.11 | 0.30 | 0.08 | 100.00 |
| | | | Std Dev | 0.22 | 0.01 | 0.09 | 0.01 | 0.01 | 0.16 | 0.10 | 0.01 | 0.09 | 0.01 | 0.01 | 0.00 |

EPMA WDS data for Fe were used for standardization for all samples. EPMA WDS data were also used in calculating weighted averages and errors for Fe, S, Ni, and Co. TEM EDS data were used in calculating the error for QUE 99177 C1S2 pyrrhotite and pentlandite.

PPI = pyrrhotite-pentlandite intergrowth, SRM = sulfide-rimmed metal, Po = pyrrhotite, Pn = pentlandite, M = metal, Std Dev = standard deviation.

[1]TEM EDS data may not agree with EPMA WDS data owing to differences in the spatial resolution of the two techniques. This is particularly true for grains with fine-scale compositional heterogeneities, such as pyrrhotite with submicron pentlandite inclusions.

**Table A2.** EPMA conditions

| Phase | EPMA Conditions | Fe | S | Ni | Co | Cr |
|---|---|---|---|---|---|---|
| Po/Pn | Crystal | LIF/H | PETJ/L | LIF/H | LIFH | PETJ/L |
| | Count Time (s) | 20–30 | 30–40 | 20–40 | 30–40 | 30–60 |
| | Detection Limit (wt.%) | 0.04–0.12 | 0.01–0.02 | 0.03–0.04 | 0.03–0.16 | 0.01–0.04 |
| | Standard | Pyrite | Pyrite | Ni metal | Co metal | Chromite |
| M | Crystal | LIFH | PETL | LIFH | LIFH | LIFH |
| | Count Time (s) | 20 | 30 | 20 | 30 | 30 |
| | Detection Limit (wt.%) | 0.04 | 0.01 | 0.04 | 0.03 | 0.02 |
| | Standard | Fe metal | Pyrite | Ni metal | Co metal | Cr metal |

Po = pyrrhotite, Pn = pentlandite, M = metal.

**Table A3.** SXRF raw data

| Sample | Phase | Filter | Spectrum | Element | Peak | Energy (keV) | FWHM | Area | Background | Area/MDL | Area/Bkg | Imprecision | Bias | Abundance (ppm) | MDL (ppm) |
|---|---|---|---|---|---|---|---|---|---|---|---|---|---|---|---|
| QUE97990 (CM2, PPI, C12S2) | Po | 110Al | spot1 | Fe | Ka | 6.4 | 0.1543 | 38796638.1 | 78321 | 46209.8 | 495.4 | (FIXED) | 61.8 | 618000 | 0.075 |
| | | | | S | – | – | – | – | – | – | – | (FIXED) | 36.77 | 367700 | – |
| | | | | Ni | Ka | 7.472 | 0.1621 | 3141771.6 | 72984 | 3876.5 | 43 | 0.012 | 0.0000001 | 12500 | 0.310 |
| | | | | Co | Kb | 7.649 | 0.1633 | 104563.7 | 69968 | 131.8 | 1.5 | 0.003347 | 0 | 3350 | 0.039 |
| | | | | Cu | Ka | 8.041 | 0.166 | 348658 | 66576 | 450.4 | 5.2 | 0.0008445 | 0 | 844.516 | 0.533 |
| | | | | Ge | Ka | 9.876 | 0.1777 | 29279.3 | 47351 | 44.9 | 0.6 | 0.000024 | 0 | 23.978 | 1.873 |
| | | | | Zn | Ka | 8.631 | 0.1699 | 23056 | 59649 | 31.5 | 0.4 | 0.0000349 | 0 | 34.901 | 0.903 |
| | | | | Se | Kb1 | 12.495 | 0.1927 | 54443.9 | 36991 | 94.4 | 1.5 | 0.0001501 | 0 | 150.102 | 0.629 |
| | | 110Al | spot2 | Fe | Ka | 6.4 | 0.1541 | 35167536.7 | 77455 | 42120.7 | 454 | (FIXED) | 61.8 | 618000 | 0.068 |
| | | | | S | – | – | – | – | – | – | – | (FIXED) | 36.77 | 367700 | – |
| | | | | Ni | Ka | 7.472 | 0.1633 | 2904511.7 | 73045 | 3582.3 | 39.8 | 0.012 | 0 | 12500 | 0.287 |
| | | | | Co | Kb | 7.649 | 0.1647 | 83937 | 70359 | 105.5 | 1.2 | 0.002891 | 0 | 2890 | 0.037 |
| | | | | Cu | Ka | 8.041 | 0.1679 | 401135.9 | 68840 | 509.6 | 5.8 | 0.0010405 | 0 | 1040 | 0.490 |
| | | | | Ge | Ka | 9.876 | 0.1818 | 26931.9 | 57476 | 37.4 | 0.5 | 0.0000233 | 0 | 23.303 | 1.605 |
| | | | | Zn | Ka | 8.631 | 0.1725 | 0 | 71288 | 0 | 0 | 0 | 0 | 0 | 0.300 [1] |
| | | | | Se | Kb1 | 12.495 | 0.1995 | 55062.5 | 43810 | 87.7 | 1.3 | 0.0001592 | 0 | 159.178 | 0.551 |
| | | 110Al | spot3 | Fe | Ka | 6.4 | 0.1532 | 19925479.1 | 50994 | 29412.2 | 390.7 | (FIXED) | 61.8 | 618000 | 0.048 |
| | | | | S | – | – | – | – | – | – | – | (FIXED) | 36.77 | 367700 | – |
| | | | | Ni | Ka | 7.472 | 0.1619 | 1635524.7 | 45935 | 2543.7 | 35.6 | 0.012 | 0 | 12400 | 0.205 |
| | | | | Co | Kb | 7.649 | 0.1633 | 49227.6 | 43502 | 78.7 | 1.1 | 0.0029922 | 0 | 2990 | 0.026 |
| | | | | Cu | Ka | 8.041 | 0.1663 | 150988.6 | 41084 | 248.3 | 3.7 | 0.0006912 | 0 | 691.22 | 0.359 |
| | | | | Ge | Ka | 9.876 | 0.1795 | 24197.2 | 39672 | 40.5 | 0.6 | 0.000037 | 0 | 36.953 | 1.096 |
| | | | | Zn | Ka | 8.631 | 0.1707 | 5082.6 | 41230 | 8.3 | 0.1 | 0.0000145 | 0 | 14.458 | 0.574 |
| | | | | Se | Kb1 | 12.495 | 0.1963 | 27495.5 | 40386 | 45.6 | 0.7 | 0.0001403 | 0 | 140.294 | 0.325 |
| | | 110Al | spot4 | Fe | Ka | 6.4 | 0.1524 | 8504889.3 | 31176 | 16056 | 272.8 | (FIXED) | 61.8 | 618000 | 0.026 |
| | | | | S | – | – | – | – | – | – | – | (FIXED) | 36.77 | 367700 | – |
| | | | | Ni | Ka | 7.472 | 0.1606 | 700846.9 | 28211 | 1390.9 | 24.8 | 0.012 | -0.0000002 | 12400 | 0.112 |
| | | | | Co | Kb | 7.649 | 0.1618 | 19189.2 | 27602 | 38.5 | 0.7 | 0.002733 | 0 | 2730 | 0.014 |
| | | | | Cu | Ka | 8.041 | 0.1646 | 128642.9 | 26655 | 262.6 | 4.8 | 0.0013794 | 0 | 1380 | 0.190 |
| | | | | Ge | Ka | 9.876 | 0.1769 | 8128.1 | 30041 | 15.6 | 0.3 | 0.0000291 | 0 | 29.085 | 0.536 |
| | | | | Zn | Ka | 8.631 | 0.1687 | 4581 | 27726 | 9.2 | 0.2 | 0.0000305 | 0 | 30.534 | 0.301 |
| | | | | Se | Kb1 | 12.495 | 0.1926 | 12355.2 | 34744 | 22.1 | 0.4 | 0.0001478 | 0 | 147.763 | 0.150 |
| | Pn | 220Al | spot2 | Fe | Ka | 6.4 | 0.1517 | 65313.8 | 15297 | 176 | 4.3 | (FIXED) | 40.42 | 404200 | 0.000 |
| | | | | S | – | – | – | – | – | – | – | (FIXED) | 33.38 | 333800 | – |
| | | | | Ni | Ka | 7.472 | 0.1619 | 158566.6 | 15780 | 420.8 | 10 | (FIXED) | 24.7 | 247000 | 0.002 |
| | | | | Co | Ka | 6.925 | 0.1568 | 4628.8 | 15721 | 12.3 | 0.3 | 0.0074565 | 0.0001835 | 7460 | 0.002 |
| | | | | Cu | Kb | 8.907 | 0.1744 | 661 | 20078 | 1.6 | 0 | 0.0005097 | 0.0000018 | 509.655 | 0.003 |
| | | | | Ge | Ka | 9.876 | 0.1823 | 0 | 23289 | 0 | 0 | 0 | 0 | 0 | 0.500 [2] |
| | | | | Zn | Ka | 8.631 | 0.1721 | 4056.2 | 19128 | 9.8 | 0.2 | 0.0004501 | 0.0000166 | 450.153 | 0.022 |
| | | | | Se | Kb1 | 12.495 | 0.2019 | 4143.8 | 31995 | 7.7 | 0.1 | 0.0003674 | 0.0000072 | 367.413 | 0.021 |
| | | 220Al | spot3 | Fe | Ka | 6.4 | 0.1527 | 56381.4 | 15487 | 151 | 3.6 | (FIXED) | 40.42 | 404200 | 0.000 |
| | | | | S | – | – | – | – | – | – | – | (FIXED) | 33.38 | 333800 | |
| | | | | Ni | Ka | 7.472 | 0.1633 | 134936.4 | 16010 | 355.5 | 8.4 | (FIXED) | 24.7 | 247000 | 0.001 |
| | | | | Co | Ka | 6.925 | 0.158 | 3849.4 | 15906 | 10.2 | 0.2 | 0.0071817 | 0.0001852 | 7180 | 0.001 |
| | | | | Cu | Kb | 8.907 | 0.1764 | 824.6 | 20228 | 1.9 | 0 | 0.0007365 | 0.000004 | 736.481 | 0.003 |
| | | | | Ge | Ka | 9.876 | 0.1846 | 0 | 23931 | 0 | 0 | 0 | 0 | 0 | 0.500 [2] |
| | | | | Zn | Ka | 8.631 | 0.1739 | 4103.4 | 19275 | 9.9 | 0.2 | 0.0005274 | 0.0000239 | 527.491 | 0.019 |
| | | | | Se | Kb1 | 12.495 | 0.205 | 4189.4 | 33423 | 7.6 | 0.1 | 0.0004302 | 0.0000103 | 430.236 | 0.018 |
| | | 250Al | spot4 | Fe | Ka | 6.4 | 0.1542 | 33928.3 | 15382 | 91.2 | 2.2 | (FIXED) | 40.42 | 404200 | 0.000 |
| | | | | S | – | – | – | – | – | – | – | (FIXED) | 33.38 | 333800 | – |
| | | | | Ni | Ka | 7.472 | 0.1632 | 72555.9 | 15705 | 193 | 4.6 | (FIXED) | 24.7 | 247000 | 0.001 |
| | | | | Co | Ka | 6.925 | 0.1587 | 2876.4 | 15732 | 7.6 | 0.2 | 0.0075893 | 0.0005925 | 7590 | 0.001 |
| | | | | Cu | Kb | 8.907 | 0.1744 | 663.6 | 20151 | 1.6 | 0 | 0.0006054 | 0.0000081 | 605.389 | 0.003 |
| | | | | Ge | Ka | 9.876 | 0.1814 | 0 | 23449 | 0 | 0 | 0 | 0 | 0 | 0.500 [2] |
| | | | | Zn | Ka | 8.631 | 0.1723 | 4119.7 | 19258 | 9.9 | 0.2 | 0.0005563 | 0.0000804 | 556.63 | 0.018 |
| | | | | Se | Kb1 | 12.495 | 0.1989 | 3451.1 | 31863 | 6.4 | 0.1 | 0.0002996 | 0.0000151 | 299.605 | 0.021 |
| QUE99177 (CR2, PPI, C1S2) | Po | 160Al | spot1 | Fe | Ka | 6.4 | 0.1536 | 25708793.4 | 72196 | 31893.6 | 356.1 | (FIXED) | 62.49 | 624900 | 0.051 |
| | | | | S | – | – | – | – | – | – | – | (FIXED) | 36.91 | 369100 | – |
| | | | | Ni | Ka | 7.472 | 0.1626 | 2374692.1 | 64207 | 3123.9 | 37 | 0.0086817 | -0.0000001 | 8680 | 0.360 |
| | | | | Co | Kb | 7.649 | 0.164 | 44758.9 | 59785 | 61 | 0.7 | 0.0012466 | 0 | 1250 | 0.049 |
| | | | | Cu | Ka | 8.041 | 0.1672 | 85682 | 57058 | 119.6 | 1.5 | 0.0001615 | 0 | 161.522 | 0.740 |
| | | | | Ge | Ka | 9.876 | 0.1808 | 5069.2 | 62394 | 6.8 | 0.1 | 0.0000024 | 0 | 2.358 | ***2.884*** |
| | | | | Zn | Ka | 8.631 | 0.1717 | 1935.2 | 58480 | 2.7 | 0 | 0.000002 | 0 | 2 | 1.350 |
| | | | | Se | Kb1 | 12.495 | 0.1982 | 36166.5 | 58448 | 49.9 | 0.6 | 0.000047 | 0 | 46.974 | 1.062 |
| | | 140Al | spot2 | Fe | Ka | 6.4 | 0.1536 | 25488859.4 | 61918 | 34144.5 | 411.7 | (FIXED) | 62.49 | 624900 | 0.055 |
| | | | | S | – | – | – | – | – | – | – | (FIXED) | 36.91 | 369100 | – |
| | | | | Ni | Ka | 7.472 | 0.1626 | 1788813.9 | 57979 | 2476.3 | 30.9 | 0.0079801 | 0.0000001 | 7980 | 0.310 |
| | | | | Co | Kb | 7.649 | 0.164 | 37123.1 | 54994 | 52.8 | 0.7 | 0.0012889 | 0 | 1290 | 0.041 |
| | | | | Cu | Ka | 8.041 | 0.1671 | 74196.7 | 53514 | 106.9 | 1.4 | 0.0001821 | 0 | 182.09 | 0.587 |
| | | | | Ge | Ka | 9.876 | 0.1808 | 10915.7 | 64708 | 14.3 | 0.2 | 0.0000075 | 0 | 7.473 | 1.914 |
| | | | | Zn | Ka | 8.631 | 0.1717 | 1842 | 56449 | 2.6 | 0 | 0.0000026 | 0 | 2.607 | 0.997 |
| | | | | Se | Kb1 | 12.495 | 0.1982 | 36307.5 | 58204 | 50.2 | 0.6 | 0.0000746 | 0 | 74.643 | 0.673 |
| | | 140Al | spot3 | Fe | Ka | 6.4 | 0.1537 | 27629805.7 | 65430 | 36005.4 | 422.3 | (FIXED) | 62.49 | 624900 | 0.058 |
| | | | | S | – | – | – | – | – | – | – | (FIXED) | 36.91 | 369100 | – |
| | | | | Ni | Ka | 7.472 | 0.1627 | 1628640 | 59770 | 2220.6 | 27.2 | 0.0067036 | 0 | 6700 | 0.331 |
| | | | | Co | Kb | 7.649 | 0.1641 | 36379.2 | 56735 | 50.9 | 0.6 | 0.0011654 | 0 | 1170 | 0.044 |
| | | | | Cu | Ka | 8.041 | 0.1672 | 107869.8 | 54582 | 153.9 | 2 | 0.0002443 | 0 | 244.279 | 0.630 |
| | | | | Ge | Ka | 9.876 | 0.1807 | 8025.4 | 66474 | 10.4 | 0.1 | 0.0000051 | 0 | 5.066 | 2.053 |
| | | | | Zn | Ka | 8.631 | 0.1717 | 1512.6 | 57065 | 2.1 | 0 | 0.000002 | 0 | 1.975 | 1.063 |
| | | | | Se | Kb1 | 12.495 | 0.198 | 39998.6 | 59753 | 54.5 | 0.7 | 0.0000759 | 0 | 75.857 | 0.718 |
| | | 130Al | spot4 | Fe | Ka | 6.4 | 0.1536 | 24951031.9 | 59761 | 34021.9 | 417.5 | (FIXED) | 62.49 | 624900 | 0.054 |
| | | | | S | – | – | – | – | – | – | – | (FIXED) | 36.91 | 369100 | – |
| | | | | Ni | Ka | 7.472 | 0.1628 | 1512763 | 57912 | 2095.4 | 26.1 | 0.0075841 | 0 | 7580 | 0.276 |
| | | | | Co | Kb | 7.649 | 0.1642 | 35337.6 | 54817 | 50.3 | 0.6 | 0.0013937 | 0 | 1390 | 0.036 |
| | | | | Cu | Ka | 8.041 | 0.1674 | 101523.1 | 53661 | 146.1 | 1.9 | 0.0002891 | 0 | 289.098 | 0.505 |
| | | | | Ge | Ka | 9.876 | 0.1812 | 8131.5 | 69583 | 10.3 | 0.1 | 0.0000069 | 0 | 6.867 | 1.500 |
| | | | | Zn | Ka | 8.631 | 0.172 | 1596.9 | 60713 | 2.2 | 0 | 0.0000027 | 0 | 2.691 | 0.818 |
| | | | | Se | Kb1 | 12.495 | 0.1988 | 37585.7 | 62227 | 50.2 | 0.6 | 0.0000989 | 0 | 98.899 | 0.508 |
| | Pn | 180Al | spot3 | Fe | Ka | 6.4 | 0.1516 | 66524.7 | 18470 | 163.2 | 3.6 | (FIXED) | 39.64 | 396400 | 0.000 |
| | | | S | – | – | – | – | – | – | – | (FIXED) | 33.76 | 337600 | – | |
| | | | Ni | Ka | 7.472 | 0.1617 | 146805.7 | 19052 | 354.5 | 7.7 | (FIXED) | 24.21 | 242100 | 0.001 | |
| | | | Co | Ka | 6.925 | 0.1566 | 4516.4 | 18970 | 10.9 | 0.2 | 0.0086723 | 0.0000972 | 8670 | 0.001 | |
| | | | Cu | Kb | 8.907 | 0.1742 | 1173.7 | 24622 | 2.5 | 0 | 0.0016617 | 0.0000074 | 1660 | 0.002 | |
| | | | Ge | Ka | 9.876 | 0.182 | 0 | 29405 | 0 | 0 | 0 | 0 | 0 | 0.500 | 3 |
| | | | Zn | Ka | 8.631 | 0.1718 | 3266.2 | 23871 | 7 | 0.1 | 0.0006408 | 0.0000128 | 640.766 | 0.011 | |
| | | | Se | Kb1 | 12.495 | 0.2015 | 4925.4 | 43644 | 7.9 | 0.1 | 0.0010339 | 0.0000216 | 1030 | 0.008 | |
| | | 250Al spot4 | Fe | Ka | 6.4 | 0.1584 | 46077.1 | 18270 | 113.6 | 2.5 | (FIXED) | 39.64 | 396400 | 0.000 | |
| | | | S | – | – | – | – | – | – | – | (FIXED) | 33.76 | 337600 | – | |
| | | | Ni | Ka | 7.472 | 0.1647 | 92447.7 | 18347 | 227.5 | 5 | (FIXED) | 24.21 | 242100 | 0.001 | |
| | | | Co | Ka | 6.925 | 0.1615 | 4423.7 | 18375 | 10.9 | 0.2 | 0.0084202 | 0.0008654 | 8420 | 0.001 | |
| | | | Cu | Kb | 8.907 | 0.1726 | 1496.1 | 24542 | 3.2 | 0.1 | 0.0009819 | 0.0000248 | 981.983 | 0.003 | |
| | | | Ge | Ka | 9.876 | 0.1775 | 0 | 29297 | 0 | 0 | 0 | 0 | 0 | 0.500 | 3 |
| | | | Zn | Ka | 8.631 | 0.1711 | 3632.5 | 23879 | 7.8 | 0.2 | 0.0003529 | 0.0000376 | 353.026 | 0.022 | |
| | | | Se | Kb1 | 12.495 | 0.1897 | 8357.8 | 43834 | 13.3 | 0.2 | 0.0005231 | 0.0000536 | 523.263 | 0.025 | |
| EET92042 (CR2, SRM, C3S3) | Po | 120Al spot1 | Fe | Ka | 6.4 | 0.1545 | 40932865.4 | 73809 | 50222.3 | 554.6 | (FIXED) | 62.74 | 627400 | 0.080 | |
| | | | S | – | – | – | – | – | – | – | (FIXED) | 35.97 | 359700 | – | |
| | | | Ni | Ka | 7.472 | 0.1637 | 2207118.7 | 65722 | 2869.8 | 33.6 | 0.0075958 | 0.0000001 | 7600 | 0.378 | |
| | | | Co | Kb | 7.649 | 0.1651 | 61747.6 | 62598 | 82.3 | 1 | 0.001687 | 0 | 1690 | 0.049 | |
| | | | Cu | Ka | 8.041 | 0.1683 | 106010.8 | 58899 | 145.6 | 1.8 | 0.000213 | 0 | 212.957 | 0.684 | |
| | | | Ge | Ka | 9.876 | 0.182 | 0 | 60053 | 0 | 0 | 0 | 0 | 0 | 2.000 | 4 |
| | | | Zn | Ka | 8.631 | 0.1728 | 4086.1 | 59537 | 5.6 | 0.1 | 0.000005 | 0 | 4.959 | 1.129 | |
| | | | Se | Kb1 | 12.495 | 0.1997 | 110440.6 | 56710 | 154.6 | 1.9 | 0.0002218 | 0 | 221.792 | 0.697 | |
| | | 120Al spot2 | Fe | Ka | 6.4 | 0.1546 | 41155375.4 | 83291 | 47534.2 | 494.1 | (FIXED) | 62.74 | 627400 | 0.076 | |
| | | | S | – | – | – | – | – | – | – | (FIXED) | 35.97 | 359700 | – | |
| | | | Ni | Ka | 7.472 | 0.164 | 2583165.6 | 76928 | 3104.5 | 33.6 | 0.0088402 | -0.0000001 | 8840 | 0.351 | |
| | | | Co | Kb | 7.649 | 0.1655 | 60482.6 | 73310 | 74.5 | 0.8 | 0.001643 | 0 | 1640 | 0.045 | |
| | | | Cu | Ka | 8.041 | 0.1688 | 276374.4 | 67913 | 353.5 | 4.1 | 0.0005522 | 0 | 552.191 | 0.640 | |
| | | | Ge | Ka | 9.876 | 0.1831 | 0 | 68837 | 0 | 0 | 0 | 0 | 0 | 2.000 | 4 |
| | | | Zn | Ka | 8.631 | 0.1735 | 2805 | 66240 | 3.6 | 0 | 0.0000034 | 0 | 3.386 | 1.063 | |
| | | | Se | Kb1 | 12.495 | 0.2013 | 115552.9 | 61477 | 155.3 | 1.9 | 0.000231 | 0 | 230.953 | 0.672 | |
| | | 120Al spot3 | Fe | Ka | 6.4 | 0.1546 | 41766318.9 | 78110 | 49814 | 534.7 | (FIXED) | 62.74 | 627400 | 0.079 | |
| | | | S | – | – | – | – | – | – | – | (FIXED) | 35.97 | 359700 | – | |
| | | | Ni | Ka | 7.472 | 0.164 | 2236346.5 | 72207 | 2774.1 | 31 | 0.0075408 | 0 | 7540 | 0.368 | |
| | | | Co | Kb | 7.649 | 0.1655 | 51887.4 | 68930 | 65.9 | 0.8 | 0.0013891 | 0 | 1390 | 0.047 | |
| | | | Cu | Ka | 8.041 | 0.1688 | 162292.5 | 66108 | 210.4 | 2.5 | 0.0003195 | 0 | 319.496 | 0.659 | |
| | | | Ge | Ka | 9.876 | 0.183 | 2801.9 | 65714 | 3.6 | 0 | 0.0000017 | 0 | 1.72 | ***2.093*** | |
| | | | Zn | Ka | 8.631 | 0.1735 | 2892.3 | 69033 | 3.7 | 0 | 0.0000034 | 0 | 3.439 | 1.076 | |
| | | | Se | Kb1 | 12.495 | 0.2011 | 120075.7 | 67801 | 153.7 | 1.8 | 0.0002364 | 0 | 236.426 | 0.650 | |
| | | 120Al spot4 | Fe | Ka | 6.4 | 0.1547 | 42811426.8 | 78777 | 50843.9 | 543.5 | (FIXED) | 62.74 | 627400 | 0.081 | |
| | | | S | – | – | – | – | – | – | – | (FIXED) | 35.97 | 359700 | – | |
| | | | Ni | Ka | 7.472 | 0.1647 | 2040374 | 70399 | 2563.3 | 29 | 0.0067127 | 0 | 6710 | 0.382 | |
| | | | Co | Kb | 7.649 | 0.1663 | 50972.7 | 67104 | 65.6 | 0.8 | 0.0013313 | 0 | 1330 | 0.049 | |
| | | | Cu | Ka | 8.041 | 0.1697 | 152105.5 | 67670 | 194.9 | 2.2 | 0.0002921 | 0 | 292.142 | 0.667 | |
| | | | Ge | Ka | 9.876 | 0.1848 | 5102.5 | 69830 | 6.4 | 0.1 | 0.0000031 | 0 | 3.055 | 2.095 | |
| | | | Zn | Ka | 8.631 | 0.1747 | 3281.5 | 68639 | 4.2 | 0 | 0.0000038 | 0 | 3.808 | 1.103 | |
| | | | Se | Kb1 | 12.495 | 0.204 | 124247.8 | 69070 | 157.6 | 1.8 | 0.0002386 | 0 | 238.552 | 0.661 | |
| | M | 120Al spot1 | Fe | Ka | 6.4 | 0.1562 | 107257575 | 137968 | 96253.7 | 777.4 | (FIXED) | 92.63 | 926300 | 0.104 | |
| | | | Ni | Ka | 7.472 | 0.1573 | 28390086.1 | 128121 | 26438.4 | 221.6 | 0.063 | 0.0000292 | 63200 | 0.418 | |
| | | | Co | Kb | 7.649 | 0.1575 | 1114386.8 | 121026 | 1067.8 | 9.2 | 0.02 | 0.0000006 | 19500 | 0.055 | |
| | | | Cu | Ka | 8.041 | 0.1579 | 698271.2 | 123665 | 661.9 | 5.6 | 0.0008904 | 0 | 890.407 | 0.743 | |
| | | | Ge | Ka | 9.876 | 0.1597 | 21745.1 | 132946 | 19.9 | 0.2 | 0.0000083 | 0 | 8.289 | 2.401 | |
| | | | Zn | Ka | 8.631 | – | – | – | – | – | – | – | 0 | 0.5 | 5 |
| | | | Se | Kb1 | 12.495 | 0.1619 | 7747.7 | 57802 | 10.7 | 0.1 | 0.0000091 | 0 | 9.109 | 1.175 | |
| | | 125Al spot2 | Fe | Ka | 6.4 | 0.1571 | 77160072.6 | 118148 | 74827 | 653.1 | (FIXED) | 92.63 | 926300 | 0.081 | |
| | | | Ni | Ka | 7.472 | 0.1654 | 21575081.2 | 115313 | 21178.3 | 187.1 | 0.064 | 0.0000304 | 63700 | 0.332 | |
| | | | Co | Kb | 7.649 | 0.1667 | 350067.8 | 112840 | 347.4 | 3.1 | 0.0080897 | 0.0000001 | 8090 | 0.043 | |
| | | | Cu | Ka | 8.041 | 0.1695 | 476272.5 | 119029 | 460.2 | 4 | 0.0007905 | 0 | 790.503 | 0.582 | |
| | | | Ge | Ka | 9.876 | 0.182 | 0 | 92996 | 0 | 0 | 0 | 0 | 0 | 2.000 | 6 |
| | | | Zn | Ka | 8.631 | – | – | – | – | – | – | – | 0 | 0.5 | 5 |
| | | | Se | Kb1 | 12.495 | 0.1978 | 4986.1 | 55064 | 7.1 | 0.1 | 0.0000073 | 0 | 7.28 | 0.975 | |
| | | 125Al spot3 | Fe | Ka | 6.4 | 0.1551 | 44364323.3 | 92323 | 48669.6 | 480.5 | (FIXED) | 92.63 | 926300 | 0.053 | |
| | | | Ni | Ka | 7.472 | 0.1637 | 12614537 | 91108 | 13930.7 | 138.5 | 0.065 | 0.0000308 | 64800 | 0.215 | |
| | | | Co | Kb | 7.649 | 0.165 | 179766.3 | 88685 | 201.2 | 2 | 0.0072262 | 0 | 7230 | 0.028 | |
| | | | Cu | Ka | 8.041 | 0.168 | 309294 | 83932 | 355.9 | 3.7 | 0.0008927 | 0 | 892.723 | 0.399 | |
| | | | Ge | Ka | 9.876 | 0.181 | 0 | 71960 | 0 | 0 | 0 | 0 | 0 | 2.000 | 6 |
| | | | Zn | Ka | 8.631 | – | – | – | – | – | – | – | 0 | 0.5 | 5 |
| | | | Se | Kb1 | 12.495 | 0.1976 | 7828.3 | 54613 | 11.2 | 0.1 | 0.0000199 | 0 | 19.882 | 0.563 | |
| | | 125Al spot4 | Fe | Ka | 6.4 | 0.1541 | 27872883.6 | 78150 | 33235 | 356.7 | (FIXED) | 92.63 | 926300 | 0.036 | |
| | | | Ni | Ka | 7.472 | 0.1633 | 7934381.2 | 76752 | 9546.6 | 103.4 | 0.065 | 0.0000293 | 64900 | 0.147 | |
| | | | Co | Kb | 7.649 | 0.1648 | 91015.9 | 76273 | 109.9 | 1.2 | 0.0058235 | 0.0000001 | 5820 | 0.019 | |
| | | | Cu | Ka | 8.041 | 0.1679 | 205926.7 | 73002 | 254.1 | 2.8 | 0.0009458 | 0 | 945.807 | 0.269 | |
| | | | Ge | Ka | 9.876 | 0.1818 | 0 | 67880 | 0 | 0 | 0 | 0 | 0 | 2.000 | 6 |
| | | | Zn | Ka | 8.631 | 0.1725 | 3644.4 | 65080 | 4.8 | 0.1 | 0.0000099 | 0 | 9.91 | 0.484 | |
| | | | Se | Kb1 | 12.495 | 0.1995 | 4093.6 | 56137 | 5.8 | 0.1 | 0.0000166 | 0 | 16.554 | 0.350 | |

FWHM = peak width (as full width as half maximum), bkg = background.

MDL = minimum detection limit, calculated as (Area/MDL)/Output; analyses below detection limit (Output < MDL) are bolded and italicized.

PPI = pyrrhotite-pentlandite intergrowth, SRM = sulfide-rimmed metal, Po = pyrrhotite, Pn = pentlandite, M = metal.

[1]DL = 0.3 from QUE97 Po spot 4, [2]DL= 0.5 from minimum of all analyses (QUE97 Po spot 4), [3]DL= 0.5 from minimum of all analyses (QUE97 Po spot 4), [4]DL = 2.0 from EET Po spot 3, [5]DL = 0.5 from EET M spot 4, [6]DL = 2.0 from EET M spot 1.

**Table A4.** Bulk *mss* calculations

| Sample | Phase | Data Type | Ni | Co | Cu | Ge | Zn | Se | Area | Density (gg/cm) | Aρ |
|---|---|---|---|---|---|---|---|---|---|---|---|
| | | | (ppm) | | | | | | | | |
| QUE97990 (CM2, PPI, C12S2) | Po | Weighted Average | 12,500 | 1,900 | 870 | 26.5 | 20.0 | 149 | 0.70 | 4.61 | 3.23 |
| | | Error[1] | 1,600 | 200 | 300 | 6.3 | 16.0 | 7.8 | | | |
| | Pn | Weighted Average | 247,000 | 7,000 | 610 | 0.3 | 500 | 360 | 0.30 | 4.8 | 1.44 |
| | | Error[1] | 910 | 300 | 110 | 0.1 | 55.0 | 65.0 | | | |
| | *Mss*[2] | Bulk | 84,855 | 3,474 | 790 | 18.4 | 168 | 214 | | | |
| | | Error[1] | 10,866 | 395 | 307 | 7.5 | 136 | 40.2 | | | |
| QUE99177 (CR2, PPI, C1S2) | Po | Weighted Average | 7,740 | 1,400 | 198 | 5.2 | 2.2 | 64.0 | 0.46 | 4.61 | 2.12 |
| | | Error[1] | 830 | 100 | 58.0 | 2.3 | 0.4 | 21.0 | | | |
| | Pn | Weighted Average | 242,100 | 8,000 | 1,110 | 0.3 | 410 | 590 | 0.54 | 4.8 | 2.59 |
| | | Error[1] | 6,600 | 200 | 480 | 0.1 | 200 | 360 | | | |
| | *Mss*[2] | Bulk | 137,943 | 5,073 | 706 | 2.5 | 229 | 356 | | | |
| | | Error[1] | 15,263 | 384 | 369 | 1.4 | 119 | 247 | | | |

PPI = pyrrhotite-pentlandite intergrowth, Po = pyrrhotite, Pn = pentlandite, *Mss* = monosulfide solid solution, Std Dev = standard deviation, Area = fraction of the area of the total grain that is the phase of interest, Aρ = area fraction times the density of the phase.

[1]For the weighted average, the error is the larger of the two between the standard deviation and the error of the average. For the bulk data, the error is the quadrature sum of the individual data (that is, pyrrhotite and pentlandite).

[2]Calculated as: $X_{po}(A\rho_{po}/A\rho_{po+pn})+X_{pn}(A\rho_{pn}/A\rho_{po+pn})$, where X is the element for which the *Mss* is being calculated.

**Table A5.** CI normalization calculations

| Sample | Phase | Data Type | Ni | Co | Cu | Ge | Zn | Se |
|---|---|---|---|---|---|---|---|---|
| | | | (ppm) | | | | | |
| QUE97990 | Po | Weighted Average | 12,450 | 1,866 | 870 | 26.5 | 20.0 | 149 |
| (CM2, PPI, | | Error[1] | 1,600 | 200 | 300 | 6.3 | 16.0 | 7.8 |
| C12S2) | | CI norm | 1.15 | 3.70 | 6.54 | 0.81 | 0.065 | 7.34 |
| | | Error[1] | 0.17 | 0.42 | 2.43 | 0.21 | 0.052 | 0.64 |
| | Pn | Weighted Average | 247,000 | 7,029 | 610 | 0.3 | 500 | 360 |
| | | Error[1] | 910 | 300 | 110 | 0.1 | 55.0 | 65.0 |
| | | CI norm | 22.64 | 13.65 | 4.59 | 0.009 | 1.62 | 17.73 |
| | | Error1 | 1.59 | 0.80 | 1.05 | 0.003 | 0.19 | 3.43 |
| | *Mss* | Bulk | 84,855 | 3,474 | 790 | 18.4 | 168 | 214 |
| | | Error[1] | 10,866 | 395 | 307 | 7.5 | 136 | 40.2 |
| | | CI norm | 7.78 | 6.77 | 5.94 | 0.56 | 0.54 | 10.55 |
| | | Error[1] | 1.14 | 0.82 | 2.46 | 0.24 | 0.44 | 2.12 |
| QUE99177 | Po | Weighted Average | 7,740 | 1,409 | 198 | 5.2 | 2.2 | 64.0 |
| (CR2, PPI, | | Error[1] | 830 | 100 | 58.0 | 2.3 | 0.4 | 21.0 |
| C1S2) | | CI norm | 0.71 | 2.73 | 1.49 | 0.16 | 0.007 | 3.15 |
| | | Error[1] | 0.09 | 0.22 | 0.48 | 0.072 | 0.001 | 1.06 |
| | Pn | Weighted Average | 242,100 | 8,037 | 1,110 | 0.3 | 410 | 590 |
| | | Error[1] | 6,600 | 200 | 480 | 0.1 | 200 | 360 |
| | | CI norm | 22.19 | 15.59 | 8.35 | 0.009 | 1.33 | 29.06 |
| | | Error[1] | 1.67 | 0.74 | 3.79 | 0.003 | 0.65 | 17.85 |
| | *Mss* | Bulk | 136,641 | 5,054 | 700 | 2.5 | 226 | 353 |
| | | Error[1] | 15,119 | 380 | 365 | 1.4 | 118 | 245 |
| | | CI norm | 12.64 | 9.89 | 5.31 | 0.077 | 0.74 | 17.56 |
| | | Error[1] | 1.66 | 0.85 | 2.87 | 0.043 | 0.39 | 12.23 |
| QUE99177 | Po | Weighted Average | 7,670 | 800 | 280 | 1.5 | 3.9 | 230.0 |
| (CR2, SRM, | | Error[1] | 880 | 200 | 150 | 1.0 | 0.7 | 7.5 |
| C3S3) | | CI norm | 0.70 | 1.56 | 2.11 | 0.046 | 0.013 | 11.33 |
| | | Error[1] | 0.09 | 0.39 | 1.17 | 0.031 | 0.002 | 0.87 |
| | M | Weighted Average | 64,100 | 3,100 | 870 | 2.8 | 2.7 | 9.8 |
| | | Error[1] | 1,000 | 100 | 65 | 3.6 | 4.9 | 6.0 |
| | | CI norm | 5.88 | 6.04 | 6.54 | 0.086 | 0.009 | 0.48 |
| | | Error[1] | 0.42 | 0.31 | 1.04 | 0.11 | 0.016 | 0.30 |
| CI chondrite[2] | | Literature value | 10,910 | 513 | 133 | 32.6 | 309 | 20 |
| | | Error | 764 | 21 | 19 | 2.9 | 12 | 1 |

PPI = pyrrhotite-pentlandite intergrowth, Po = pyrrhotite, Pn = pentlandite, *Mss* = monosulfide solid solution, M = metal, CI norm = normalized to CI chondrite.

[1]For the weighted average, the error is the larger of the two between the standard deviation and the error of the average. For the bulk and CI norm data, the error is the quadrature sum of the individual data.

[2]Data are from Palme et al. (2014), including errors.

# Appendix B to Trace elemental behavior in the solar nebula: Synchrotron X-ray fluorescence analyses of CM and CR chondritic iron sulfides and associated metal

S. A. Singerling[1]* S. R. Sutton[2, 3], A. Lanzirotti[3], M. Newville[3], and A. J. Brearley[1]
[1]Department of Earth and& Planetary Sciences, MSC-03 2040 1 University of New Mexico, Albuquerque, NM 87131, USA
[2]Department of Geophysical Sciences, University of Chicago, Chicago, IL 60637, USA
[3]Center for Advanced Radiation Sources, University of Chicago, Chicago, IL 60637, USA
*Corresponding author email: sheryl.singerling.ctr@nrl.navy.mil

## Weighted Average and Error Calculations

- Computed the 1 sigma error in net peak area (total peak–background) based on counting statistics: $\sqrt{(net + bkgd)^2 + bkgd^2}$
- Assumed the SNRLXRF errors in sensitivity ratios to Fe are ±5% based on Lu et al. (1989) and Shearer et al. (2012).
- Assumed sulfide thickness (g/cm$^2$) errors were negligible. These are FIB sections with thicknesses determined accurately by electron microscopy.
- In general, errors were computed as quadrature sums. For addition or subtraction of numbers, the error is the square root of the sum of each error squared. For multiplication and division, the error is the square root of the sum of each fractional error squared.
- Abundances: Computed the error as the quadrature sum of the peak errors and 5% (from SNRLXRF). Computed the % errors in abundances. EMPA errors come from actual EPMA analyses (Fe, S, Ni, and Co). For errors of EMPA data with only one data point (i.e., QUE 99177 pyrrhotite and pentlandite), we used the standard deviation of the SXRF analyses rounded up to the nearest hundred.
- CI-normalized abundances: Computed the errors as the quadrature sum of the % abundance errors and the % CI uncertainties (from Palme et al., 2014 table 3).
- Calculated the abundance averages as weighted averages and computed the errors as quadrature sums of the individual errors.
    - The weighted average is (the sum of each value times its weight) divided by (the sum of the weights). For weights, used the inverse of the variance (1/sigma$^2$) so values with smaller errors receive greater weight.

$$\bar{x} = \frac{\sum_{i=1}^{n}\left(\frac{x_i}{\sigma_i^2}\right)}{\sum_{i=1}^{n}\frac{1}{\sigma_i^2}},$$

$$W = \frac{\sum_{i=1}^{n} w_i X_i}{\sum_{i=1}^{n} w_i}$$

$W$ = weighted average
$n$ = number of terms to be averaged
$w_i$ = weights applied to x values
$X_i$ = data values to be averaged

and the *standard error of the weighted mean*

$$\sigma_{\bar{x}} = \sqrt{\frac{1}{\sum_{i=1}^{n}\sigma_i^{-2}}},$$

- For averages with limits, computed two unweighted averages, one assuming the limit values are the actual values and one assuming the limit values are zero. The true average must lie between these two so took the average of these two as the abundance average. For the error, assumed the difference between the two averages corresponded to the 95% confidence interval and used ¼ of this as sigma. In these cases, standard deviation is a more useful indicator of the population spread.
  - As an example, if one value is 100 and the other value is below the detection limit of 10, then the actual value of the undetectable measurement must lie in the range of 0 to 10. Therefore, the true average must lie between the average of 100 and 0 and average of 100 and 10 (i.e., 50 to 55). The methodology used here would describe this average result as 52.5 ± 1.3 (1 σ) where the uncertainty of 1.3 is obtained by assuming the total range (50 to 55) is a 4σ 95% confidence level.
- Used the larger of the two errors between the standard deviations and the error of the average. Data groups with large standard deviations suggest it may not be appropriate to average the values.

## Representative Grains

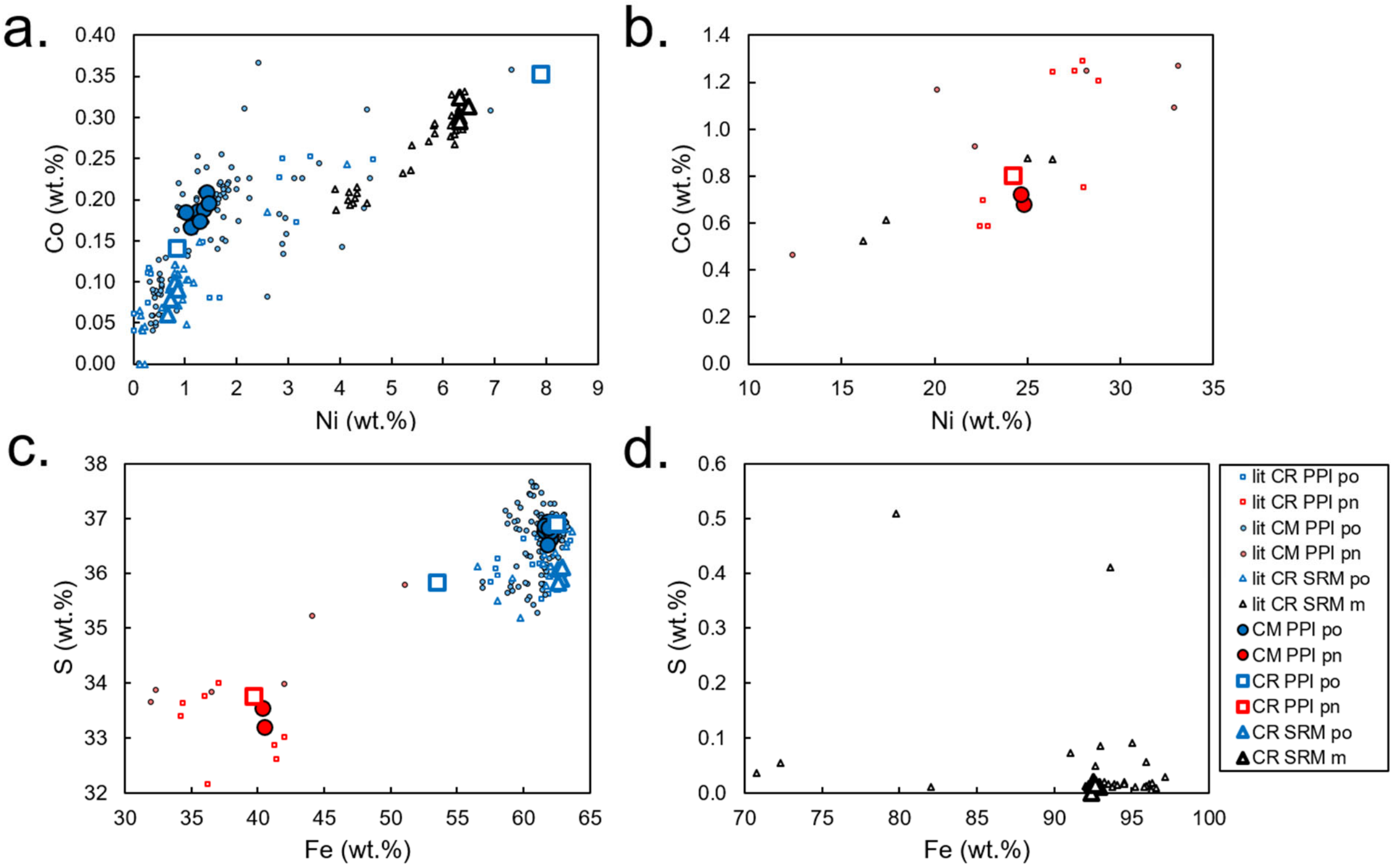

**Figure A1.** Element-element plots (Co, Ni, S, Fe in wt.%) of individual spot EPMA analyses of the FIB sections featured in this study and literature values (labeled "lit…" in the legend) from Singerling and Brearley (2018). (a–b) show Co versus Ni and (c–d) show S versus Fe. Although the sulfide and metal compositions can vary, the grains selected for SXRF analyses are representative of the overall grain populations—CM PPI and CR PPI and CR SRM. Symbols:

CM = closed, CR = open, SRM = triangles, CM PPI = circles, CR PPI = squares, Fe,Ni metal = black, pyrrhotite = blue, pentlandite = red.

## References

Lu F.-Q., Smith J. V., Sutton S. R., Rivers M. L., and Davis A. M. (1989) Synchrotron X-ray fluorescence analysis of rock-forming minerals. *Chem. Geol.* **75**, 123–143.

Shearer C. K., Burger P. V., Guan Y., Papike J. J., Sutton S. R., and Atudorei N.-V. (2012) Origin of sulfide replacement textures in lunar breccias: Implications for vapor element transport in the lunar crust. *Geochim. Cosmochim. Acta* **83**, 138-158.

Singerling S. A. and Brearley A. J. (2018) Primary iron sulfides in CM and CR carbonaceous chondrites: Insights into nebular processes. *Meteor. Planet. Sci.* **53**, 2078–2106.